\documentclass[twocolumn]{aastex62}

\pdfoutput=1 
\usepackage{amsmath,amstext,amssymb}
\usepackage[T1]{fontenc}
\usepackage[figure,figure*]{hypcap}
\usepackage{longtable}
\usepackage{booktabs}

\graphicspath{{./}{figures/}}

\received{?}
\revised{?}
\accepted{?}

\submitjournal{AAS}

\shorttitle{Amplifying Resonant Repulsion}
\shortauthors{Xu and Dai}

\begin{document}

\title{Amplifying Resonant Repulsion with Inflated Young Planets, Overlooked Inner Planets, and Non-zero Initial $\Delta$}

\correspondingauthor{Yuancheng Xu}
\email{yuancheng.xu@some.ox.ac.uk}
\author{Yuancheng Xu}
\affiliation{Department of Physics, Oxford University, Oxford, OX1 3RH, UK}

\author[0000-0002-8958-0683]{Fei Dai}
\affiliation{Institute for Astronomy, University of Hawai`i, 2680 Woodlawn Drive, Honolulu, HI 96822, USA}

\begin{abstract}
\noindent
Most multi-planet systems around mature ($\sim 5$-Gyr-old) host stars are non-resonant. Even the near-resonant planet pairs still display 1-2\% positive deviation from perfect period commensurabilities ($\Delta$) near first-order mean motion resonances (MMR). Resonant repulsion due to eccentricity tides was one of the first mechanisms proposed to explain the observed positive $\Delta$. However, the inferred rates of tidal dissipation are often implausibly rapid (with a reduced tidal quality factor $Q_p^\prime \lesssim 10$).  In this work, we attempt to amplify eccentricity tides with three previously ignored effects. 1) Planets tend to be inflated when they were younger. 2) Kepler-like Planets likely form as resonant chains parked at the disk inner edge, overlooked inner planets could have contributed to tidal dissipation of the whole system. 3) Disk migration captures planets into first-order MMR with non-zero initial deviation $\Delta$, thereby lowering the amount of dissipation needed. We show that even after accounting for all three effects, $Q_p^\prime$ can only be amplified by about one order of magnitude, and still falls short of $Q_p^\prime$ values of Solar System planets. Therefore, eccentricity tides alone cannot fully explain the observed $\Delta$ distribution. Other effects such as obliquity tides, planetesimal scattering, expanding disk inner edge, disk turbulence, divergent encounter, and dynamical instabilities must have contributed to dislodging planets from first-order MMR.

\end{abstract}

\section{Introduction}\label{sec:introduction}
There is mounting evidence that most Kepler-like planets initially formed as a resonant chain  \citep{MillsNature,Wittrock,David2019,Wood,Vach,Hamer2024,Barber,dai2024prevalenceresonanceyoungclosein,Luque} parked at the disk inner edge \citep{Masset_2006,Izidoro,Ogihara2018,Dai1136,Wong2024}. This is before the resonant configurations are disrupted during subsequent dynamical evolution \citep{Emsenhuber2021,Pichierri2020,Matsumoto,Wang2023, Goldberg_stability}. Type I disk migration, a gravitational interaction between planets and the protoplanetary disk, is likely responsible for the migration towards the central star and the capture resonant states \citep{Goldreich1979, Ward, Lin1986, McNeil, Terquem_2007, Nelson2018}.

Recently, \citet{dai2024prevalenceresonanceyoungclosein} showed that resonant configuration is predominant for planetary systems younger than 100 Myr old. On the other hand, the mature Kepler planets \citep[median age 4.6-Gyr-old, ][]{Berger_age} are generally non-resonant \citep{Borucki+2011,Fabrycky2014}. Only $15\%$ of neighboring pairs are near first-order mean-motion resonance \citep[MMR, ][]{Huang_Ormel,dai2024prevalenceresonanceyoungclosein}; and these near-resonant pairs still show $1-2\%$ positive deviation from the perfect integer period ratios. The deviation \(\Delta\), is defined as \(\frac{P_{\text{out}}/P_{\text{in}}}{(k+1)/k} - 1\), where \(P_{\text{out}}\) and \(P_{\text{in}}\) represent the orbital periods of the outer and inner planets of a neighboring pair and $k$ is the small integer specifying the specific resonance.

Resonant repulsion, particularly that due to eccentricity tides, was one of the first proposed explanations for the observed positive \(\Delta\) \citep{Papaloizou2010, Lithwick_repulsion, Batygin_repulsion, Delisle2014}. If a planet is on an eccentric orbit, the planet is strongly deformed at the pericenter and less so at the apocenter. This variation in deformation as a function of orbital phase leads to tidal dissipation within the planet, thus draining orbital energy. On the other hand, a non-zero eccentricity can be maintained by the resonant interaction between the planets near MMR. In fact, Io's strong volcanism was predicted to be maintained by the competing effect of resonant interaction and tidal dissipation \citep{Peale1976}.

While energy is dissipated, there is generally no net torque during eccentricity tidal evolution \citep{Murray}. \citet{Lithwick_repulsion} derived that the rate of change of $\Delta$ follows:
\begin{equation}
\begin{aligned}
& \  \Delta^2\frac{d\Delta}{dt} \approx 0.006^3 \left( \frac{Q}{10} \right)^{-1} \left (\frac{k_2}{0.1}\right) \left( \frac{M_{\rm p}}{10M_\oplus} \right) \left( \frac{R_{\rm p}}{2R_\oplus} \right)^{5}\\
 &\quad \times \left( \frac{M_\star}{M_\odot} \right)^{-8} \left( \frac{P_{\rm orb}}{5 \,{\rm days}} \right)^{-13/3} \left( \frac{1}{5 \,{\rm Gyr}} \right)\\
 &\quad \times (2\beta+2\beta^2) \label{eq:1}
\end{aligned}
\end{equation}

\noindent where \(k_2\), \(M_p\), \(R_p\), and \(P_{\rm orb}\) are the tidal Love number, mass, radius, and orbital period of the inner planet of resonant pair. \(\beta = \frac{M_{\text{out}} \sqrt{a_{\text{out}}}}{M_{\text{in}} \sqrt{a_{\text{in}}}}\) is the ratio of the masses and semi-major axis of the outer and inner planets engaged in resonance. The reduced tidal quality factor of inner planet is given by \(Q_p^\prime \equiv Q / k_2\).

By integrating Equation \ref{eq:1}, one can trace out the evolution of $\Delta$ from some initial value to a final value: \(\Delta_f^3 - \Delta_i^3 = \Delta_{\text{mig}}^3\). The observed $\Delta = \Delta_f$, coupled with the system's age and various planetary and stellar parameters in Equation \ref{eq:1}, could provide a constraint on the reduced tidal quality factor \(Q_p^\prime\). Previous works \citep{Lithwick_repulsion,LeeMH,Silburt,Millholland_obliquity} pointed out that the inferred $Q_p^\prime$ is often implausibly small ($Q_p^\prime \lesssim 10$). In contrast, the terretrial planets in the Solar System have $Q_p^\prime \sim 1000$ \citep{Goldreich_Soter1966}, while the giant planets have  $Q_p^\prime \gtrsim 10^5$ \citep{Tittemore,Murray}. Earth's \(Q_p^\prime\) is approximately 100 because we have a shallow ocean and thus strong dissipation due to tides breaking in coastal regions \citep{Murray,Yoder}. This is unlikely to be the case for exoplanets, especially considering that mini-Neptunes are likely covered in a thick layer of volatile envelopes \citep[e.g.][]{Yanqin2005}.

In this study, we make one more attempt to salvage the resonant repulsion due to eccentricity tides. We explore three previously ignored effects:

\begin{itemize}
    \item When planets were younger, they have inflated radii \citep[e.g.][]{Fortney}. As shown in Equation \ref{eq:1}, the instantaneous rate of change $\Delta$ goes as the fifth power of the planetary radii. A young, inflated planet should have experienced stronger eccentricity tides.

    \item Kepler-like planets likely formed initially as a chain of resonances parked at the disk inner edge \citep{Izidoro,dai2024prevalenceresonanceyoungclosein,Wong2024}. Moreover, for a chain of resonant planets, the tidal dissipation on the innermost planet could induce resonant repulsion on all planets engaged in the resonant chain \citep{Papaloizou2017,Brasser2021,Dai1136}. An overlooked inner planet could produce stronger resonant repulsion than what is inferred based on the longer-period planets in the same system.

    \item During disk migration, planets are captured into resonance with a non-zero $\Delta$ (deviation from perfect period ratio) to begin with \citep[see e.g.][]{Choksi2020,Dai1136}. The initial $\Delta_i$ depends on the ratio between semi-major axis and eccentricity damping timescales. Previous investigations on resonant repulsion have assumed $\Delta_i = 0$. This would again overestimate the amount tidal dissipation needed to reproduce the observed $\Delta_f$.
\end{itemize}

In this study, we carefully examine all three effects and determine if they can significantly lower the tidal dissipation rate (equivalent to increasing $Q_p^\prime$) required to reproduce the observed $\Delta$ distribution. We try to reconcile the $Q_p^\prime$ based on the observed near-resonant exoplanets and $Q_p^\prime$ inferred on Solar System planets. In Section 2, we describe the sample used in this study. In Section 3, we describe each of the three effects and their influences on  $Q_p^\prime$. We summarize and discuss our findings in Section 4.

\begin{figure*}
\center
\includegraphics[width = 1.\columnwidth]
{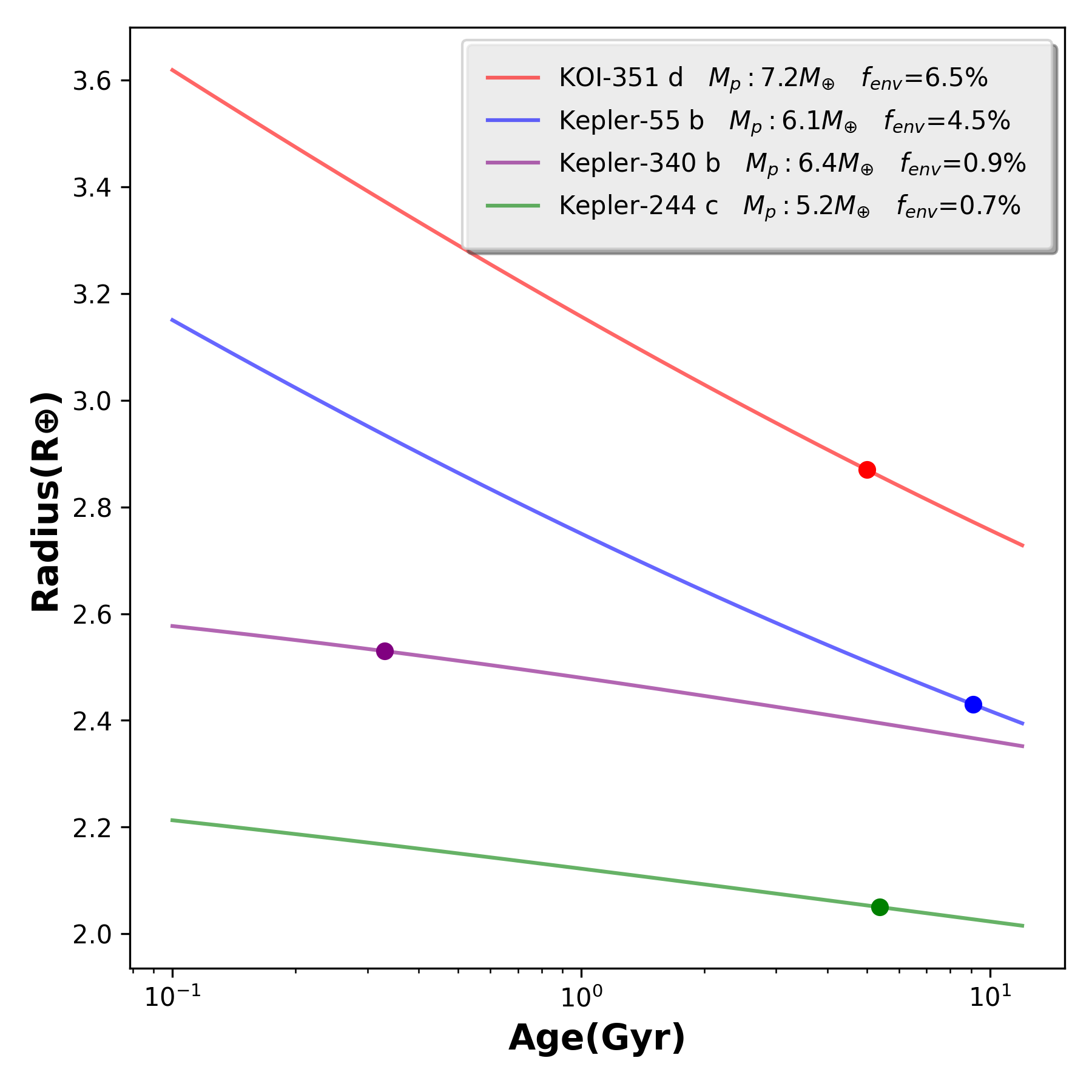}
\includegraphics[width = 1.\columnwidth]
{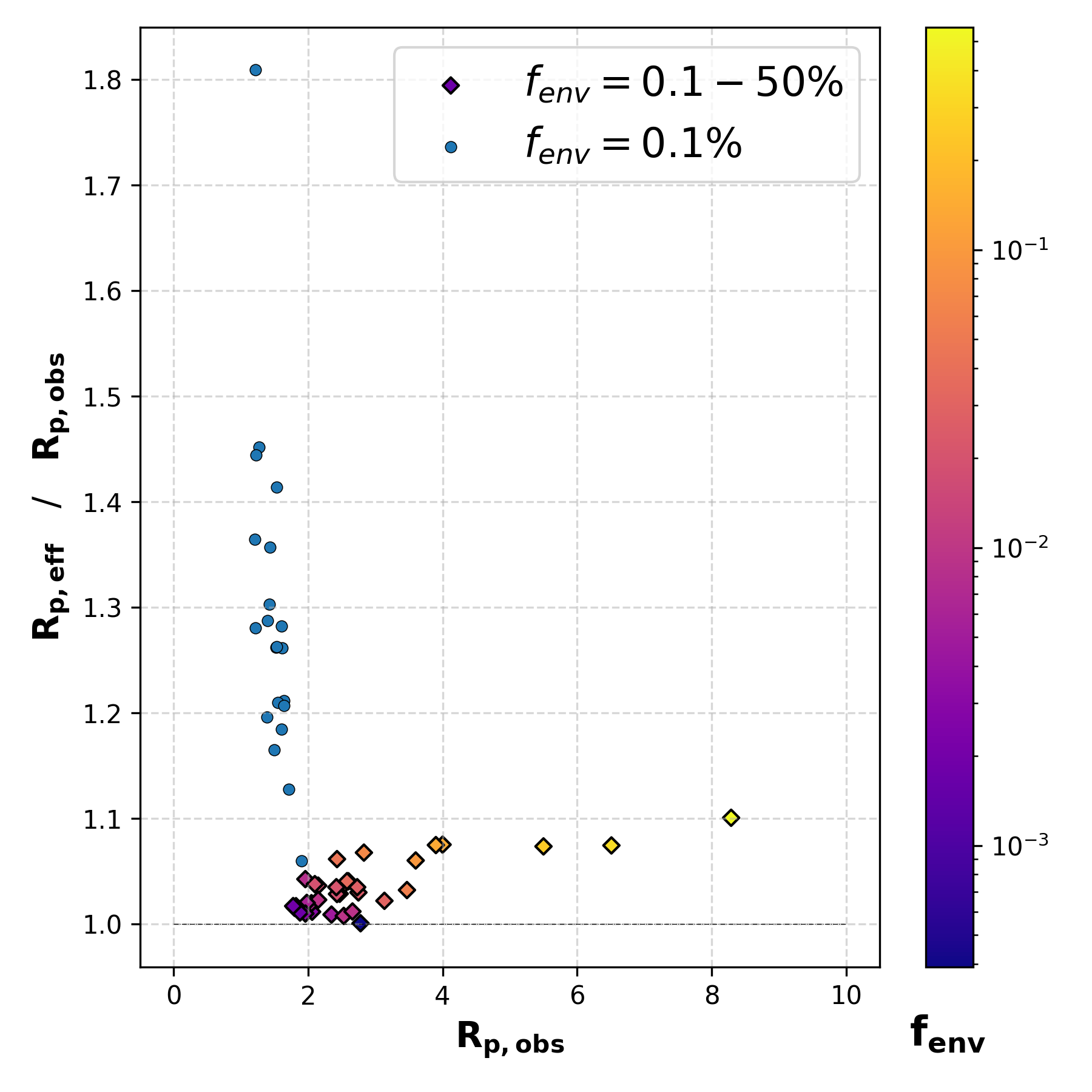}
\caption{{{\bf Left:} The radius evolution of a few planets in our sample according to fitting shown in Table 4 of (valid from 0.1-10 Gyr) \citet{Chen_2016} at a wide range of envelope mass fraction $f_{\rm env}$}. The circles are the current radii of the planets. {\bf Right:} The effective radii versus the current radii of the planets in our sample. The effective radius is a constant radius that should be used in Equation \ref{eq:1} to account for the effect of radius evolution. Super-Earth (blue points) were assumed to have an initial H/He envelope of $f_{\rm env} = 0.1\%$. Mini-Neptunes have $f_{\rm env} = 0.1-50\%$ that were solved using their current masses and radii.}\label{fig:radius}
\end{figure*}

\begin{figure*}
\center
\includegraphics[width = 1.\columnwidth]
{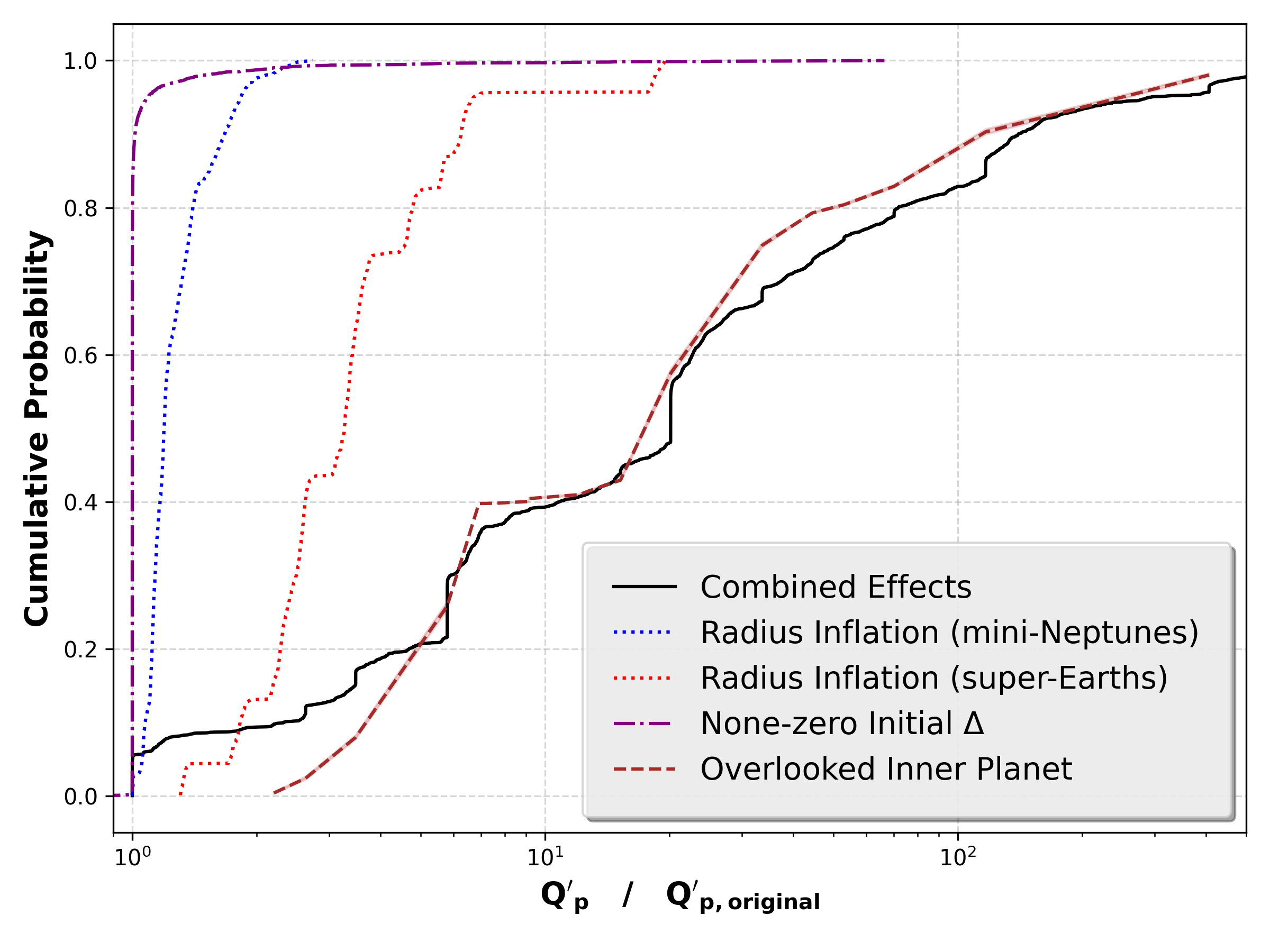}
\includegraphics[width = 1.\columnwidth]
{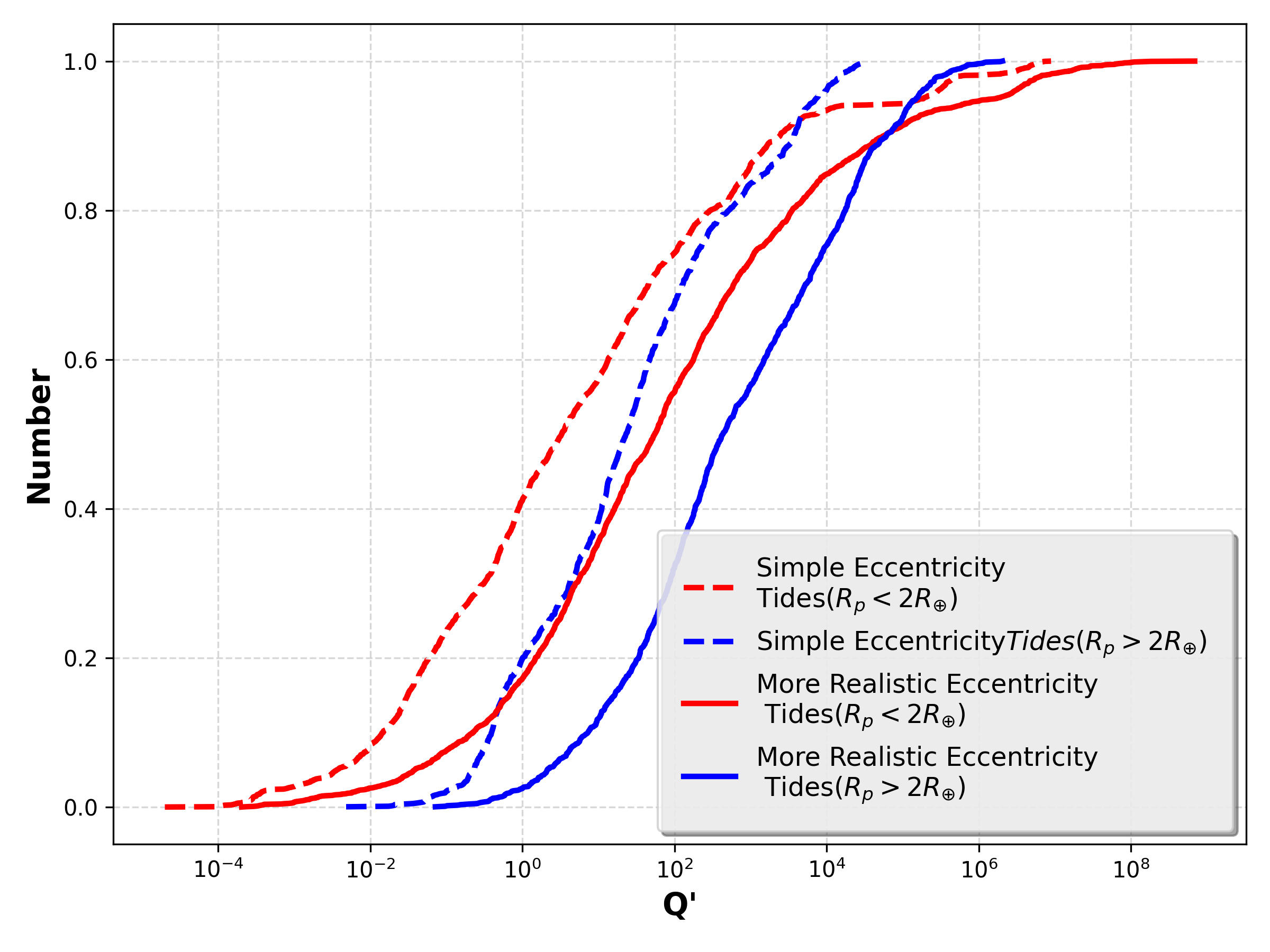}
\caption{ {\bf Left:} The change of the reduced tidal quality factor after accounting for radius evolution (dotted line), overlooked inner planets (dashed line), and non-zero initial $\Delta$ (dash-dotted line). The combined effect is shown in black solid line. Radius evolution could increase  \(log_{10}Q_p^\prime\) by 0.11$\pm$0.08 for mini-Neptuens ($>2R_\oplus$), or 0.5$\pm$0.2 for super-Earths ($<2R_\oplus$). An overlooked inner planet can increase \(log_{10}Q_p^\prime\) by $1.0 \pm0.4$. Non-zero initial $\Delta$ only increases \(log_{10}Q_p^\prime\) by no more than $<0.1$ in 95\% of the cases. The combined effect may change \(Q_p^\prime\) by one order of magnitude or so. {\bf Right:} \(Q_p^\prime\) as inferred in a simple eccentricity tide model v.s. that after accounting for radius inflation, overlooked inner planets, and non-zero initial $\Delta$. Using a simple eccentricity tide model, the tidal quality factor required to reproduce observed planetary systems are: $log_{10}(Q_p^\prime)=0.1\pm2.1$ for super-Earths (red dotted line ) and $log_{10}(Q_p^\prime)=1.5\pm1.6$ for mini-Neptunes ($>2R_\oplus$) (blue dotted line). After accounting for the three effects mentioned above, \(Q_p^\prime\) increased by about one order of magnitude: $log_{10}(Q_p^\prime)=1.2\pm2.1$ for the super-Earths and $log_{10}(Q_p^\prime)=2.4\pm1.7$ for mini-Neptunes. These values are still too small compared to Solar System values (10$^3$ for terrestrial planets, and 10$^5$ for the icy giants). }\label{fig:4}
\end{figure*}

\section{SAMPLE SELECTION}
We start with the confirmed planet sample from the NASA Exoplanet Archive \footnote{\url{https://exoplanetarchive.ipac.caltech.edu} \\ \citep{https://doi.org/10.26133/nea12}} retrieved on Aug 15 2024. We focused on the transiting planets from the Kepler sample for two reasons: 1) orbital periods are much better measured in transit surveys compared to other methods; 2) we could constrain the ages of the systems by cross-matching with the isochronal ages from \citet{Berger_age}. For planets without reported masses, we predict the mass using the observed radius and the mass-radius relationship in \citet{Otegi_2020}.

The distribution of period ratios of our sample is similar to that in previous works \citep{Fabrycky2014,Huang_Ormel,dai2024prevalenceresonanceyoungclosein}. Namely, \(\Delta\) shows a preference for positive deviation at a few percent level (see also \citealt{louden2023tidaldissipationregimesshortperiod}). We identified the near-resonant neighboring pairs as those with \(\Delta\) values within \(0-3\%\) of each first-order period commensurability (2:1, 3:2, 4:3, 5:4, and 6:5), resulting in approximately 110 pairs within this range. The choice of 3\% is based on the suggestions by previous works such as \citet{Huang_Ormel} and \cite{dai2024prevalenceresonanceyoungclosein}. Both works reported that about 15\% of neighboring pairs of confirmed planets have period ratios within this threshold. Since we are interested in the tidal dissipation of Kepler-like planets, we removed giant planets ($>10R_\oplus$). We have also removed planets whose orbital periods are too far away to have been affected by tides ($P_{\rm orb}>30$ days orbit). We have tested that the locations of the radius and period cuts do not change the qualitative result of this paper. These cuts resulted in 80 pairs of near-resonant planets in our sample (Table \ref{tab}).

\section{Methods and Results}

The key analysis in this paper is as follows. For each observed near-resonant pairs of planets (Table \ref{tab}), we numerically integrated Equation \ref{eq:1} using the observed system parameters, and the isochronal ages from \citet{Berger_age}. We can thus turn the observed period ratio deviation $\Delta$ into a constraint on the tidal dissipation rate $Q_p^\prime$. We then repeated the analysis incorporating sequentially the effects of inflated planet radii, overlooked inner planets, and non-zero initial $\Delta_i$, and observed how the inference on $Q_p^\prime$ is affected. Because there are substantial uncertainties on a number of system parameters, notably the masses of the planets, the age of the system, and other parameters that will be described shortly, we performed a bootstrap analysis by resampling 100 times with measurement uncertainties for each resonant pair in our sample.

\subsection{Radius Inflation}
Young planets with a gaseous envelope cool and contract as a function of age \citep[e.g.][]{Fortney}. Early on in a system's history, the mini-Neptunes ($>2R_\oplus$), i.e. those with H/He envelopes, should be inflated. Indeed, the majority of young planets discovered so far have significantly larger radii than the amture planets. Many young planets have radii between 4-10 $R_\oplus$  \citep[e.g.][]{David2019,Plavchan,Mann1227,Vach} i.e. substantially larger than the mature mini-Neptunes \citep[whose occurrence drops off quickly beyond 3.5$R_\oplus$,][]{Fulton}. Although most super-Earths ($<2R_\oplus$) are currently exposed rocky cores \citep{Rogers,Dressing,Dai2019} or enshrouded in heavy-mean-molecular-weight atmospheres \citep{Hu2024}, early on they could have a primordial H/He envelope that were subsequently lost due to atmospheric erosion \citep{OwenWu,Ginzburg}.

In either case,  Equation \ref{eq:1} should accounts for the inflated radii of young planets. The influence of radius evolution could be substantial since the instantaneous rate of change $\Delta$ goes as the fifth power of the planetary radii. We adopt the radius evolution model of \citet{Chen_2016} which is based on modified Experiments in Stellar Astrophysics (MESA) simulations \citep{Paxton2015}.

The model by \citet{Chen_2016} accounts for the core luminosity, heavy elements, atmospheric boundary condition, and hydrodynamic atmospheric erosion, all of which are relevant effects for the radius evolution of a planet with a H/He envelope. Using the age, the planet mass, and insolation, we solved for the current mass fraction of the H/He envelope $f_{\rm env}$ on the planets in our sample using \citet{Chen_2016}. This results in $f_{\rm env}$ between $\sim$0.1\% to $\sim$10\% for the mini-Neptunes ($>2R_\oplus$) in our sample (see also Table \ref{tab} ). For the super-Earths within our sample ($<2R_\oplus$), \citet{Chen_2016} usually predicts a vanishing low current-day $f_{\rm env}<10^{-5}$. We adopted a simple prescription that planets between 1.2-2$R_\oplus$ have an initial $f_{\rm env}=0.1\%$ whereas planets $<1.2R_\oplus$ have no primordial H/He envelope. This choice is motivated by both theoretical studies of gas accretion onto planets of different core masses \citep[e.g.][]{Lee2019} and observational work that revealed bare rocky surfaces on terrestrial-sized planets \citep[e.g.][]{Kreidberg2019,Crossfield2022,Zhang2024}.

 The left panel of Fig. \ref{fig:radius} shows the radius evolution up to 10-Gyr-old. We show a representative set of four planets that have current-day $f_{\rm env}$ between 0.2-25\%. We plugged the radius evolution into Equation \ref{eq:1} to determine how much effect it has on the inferred tidal dissipation rate $Q_p^\prime$. To facilitate our discussion, we defined an effective radius which is basically a constant radius that would produce the same amount of resonant repulsion as if the full radius evolution has been included. This effective radius is simply the time average of the fifth power of the radius $\int_{0}^{t_{age}} R(t)^5 dt=R_{\rm eff}^5\times t_{age}$.

 Using the radius evolution from \citealt{Chen_2016}, we found that the effective radius $R_{\rm eff}$ is only of order $5\pm 3 \%$ larger than the current radius for the mini-Neptunes ($>2 R_\oplus$). See Fig. \ref{fig:radius} right panel. The effect is stronger for the super-Earths ($<2 R_\oplus$), and the effective radius $R_{\rm eff}$ is about $26\pm 15 \%$ larger than the current radius. This is likely because the lower surface gravity on super-Earths cannot hold on the atmospheres as well. For terrestrial-sized planets ($<1.2 R_\oplus$), there is no radius inflation effect as we assumed no primordial envelopes.

The effect of radius inflation increased the reduced tidal quality factor log $(Q_p^\prime)$ by 0.11$\pm$0.08 for mini-Neptunes ($>2R_\oplus$), or 0.5$\pm$0.2 for super-Earths ($<2R_\oplus$) as shown in Fig. \ref{fig:4}. In other words, the effect is roughly half an order of magnitude.

\subsection{Overlooked Inner Planets}

\begin{figure*}
\center
\includegraphics[width = 2.\columnwidth]{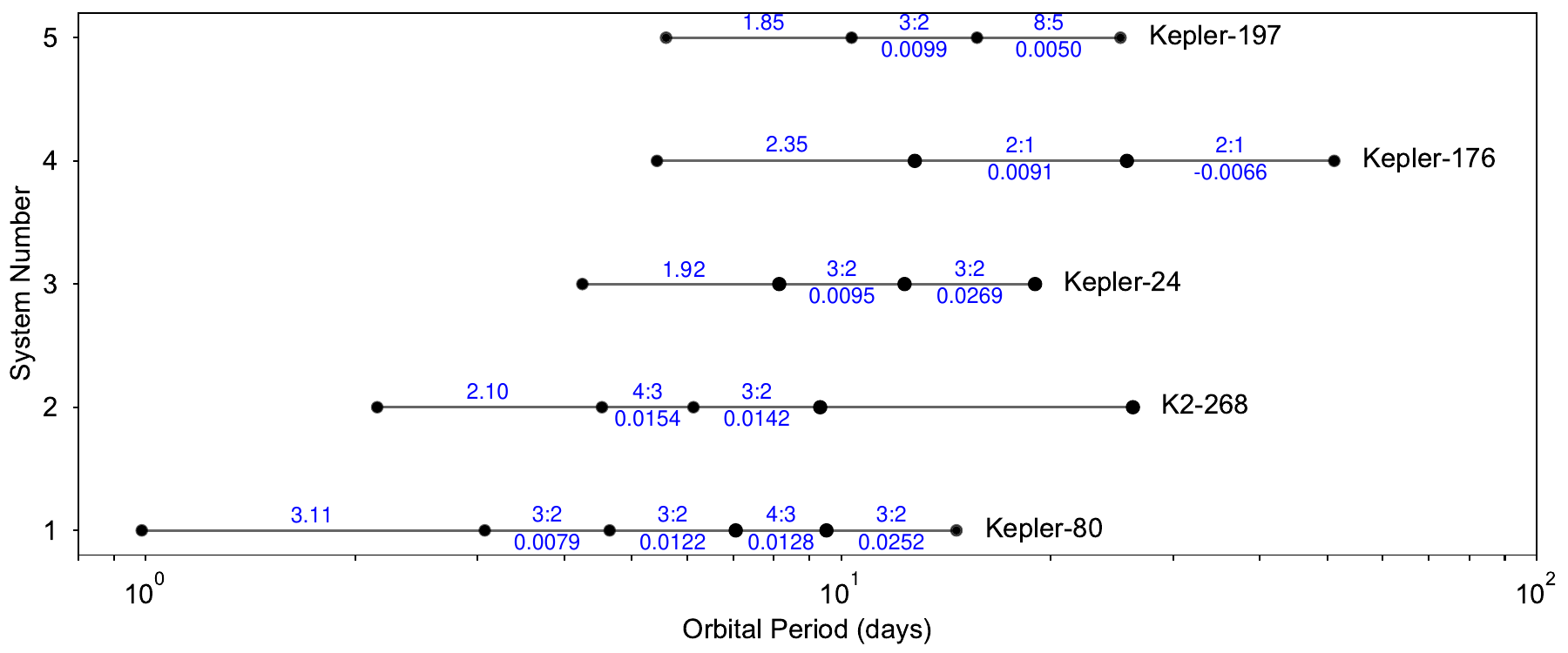}
\caption{The orbital architecture of representative near-resonant multi-planet systems where the innermost planets are currently far away from MMR. However, the innermost planets could have been part of the resonant chain initially and contributed to the resonant repulsion of the whole system. Near-resonant pairs are labeled in blue using both the MMR and the observed $\Delta$.}
\label{fig:architecture}
\end{figure*}

\begin{figure*}
\center
\includegraphics[width = 1.\columnwidth]
{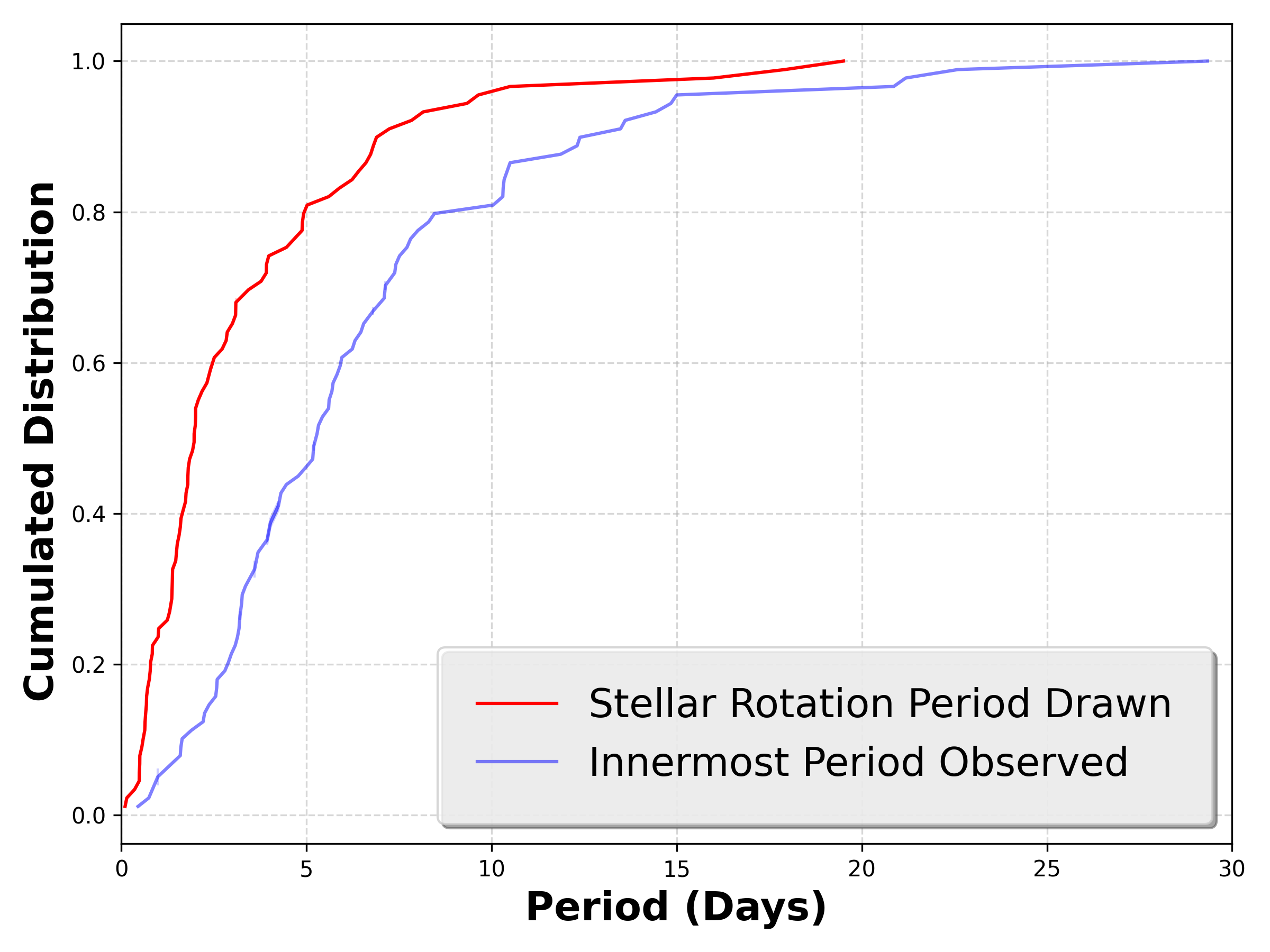}
\caption{The orbital periods of inner planets of the whole system and stellar rotation period distribution of rho Ophiuchus \citep{Rebull2018} which serves a proxy for the orbital period of an overlooked inner planet.}\label{fig:2，1}
\end{figure*}

\begin{figure*}
\center
\includegraphics[width = 1.\columnwidth]
{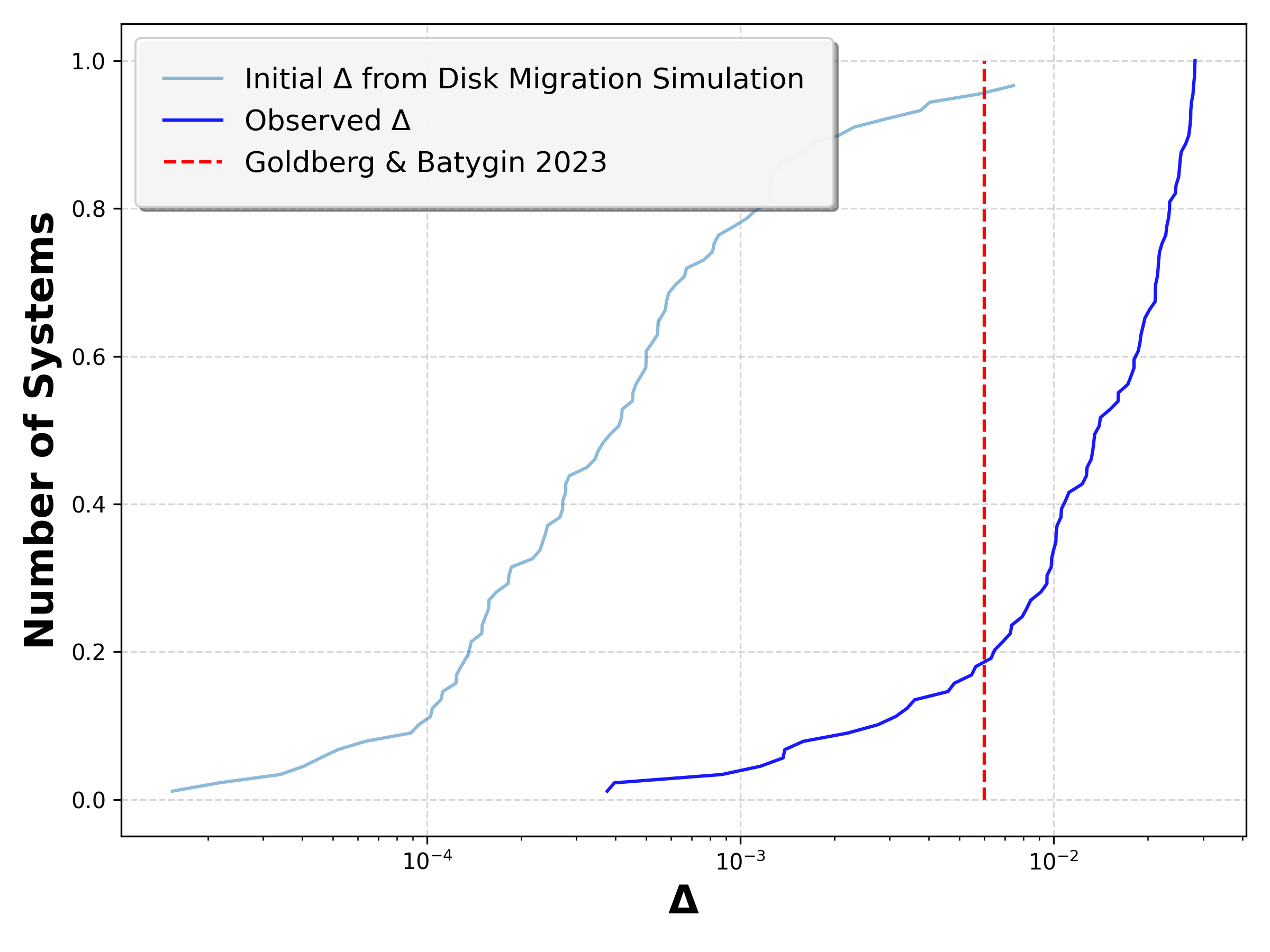}
\caption{The distribution of $\Delta_i$ ( $log_{10} (\Delta_i)=-3.4\pm0.5$) from our disk migration simulations (Keller \& Dai, in prep) and the observed $\Delta$ ($log_{10} (\Delta)=-1.9\pm0.4$) on near-resonant planets. \citet{Goldberg2023} proposed an empirical boundary for separating the librating v.s. circulating planets at $\Delta=$ 0.6\% or $log_{10} (\Delta) = -2.2$ (red dashed line). Most observed multi-planet systems are near resonance (circulating) rather in resonance (librating). Conversely, most planetary systems produced by convergent disk migration are in resonance (librating). }\label{fig:2，2}
\end{figure*}

 Kepler-like planets might have formed initially in a chain of MMR \citep{Izidoro,dai2024prevalenceresonanceyoungclosein} parked at the inner disk edge \citep{Masset_2006,Wong2024}. In addition, it has been shown both theoretically and by direct N-body simulations that the tidal dissipation on one planet could induce resonant repulsion on all other planets engaged in a resonant chain \citep{Papaloizou2017,Brasser2021,Dai1136}. This is because the resonant interaction can transmit the effect of tidal dissipation from one planet to its resonant neighbors.

 This is particularly important for systems that have planets closer to the host star than the resonant pairs of planets under investigation. These inner planets could be a planet that has not been detected yet in a transit survey, or a planet that is detected in transit but is no longer near resonance due extended resonant repulsion. In Fig \ref{fig:architecture}, we show the observed orbital architectures of Kepler-80 \citep{MacDonald2016}, Kepler-176 \citep{Rowe2014}, Kepler-197 \citep{HaddenLithwick2014}, Kepler-24 \citep{Ford2012}, and K2-268 \citep{Livingston2018}. In all of these cases, the innermost planets are currently far from MMR, but could have started as part of the resonant chain and contributed significantly to resonant repulsion before dislodging itself from the resonant chain.

We performed the following analysis to quantify how an overlooked inner planet would change our inference on the tidal dissipation rate $Q_p^\prime$. We assumed that all resonant systems were parked by inner disk edges \citep{Masset_2006,Wong2024}. Thus,we expect orbital period of the innermost planet should follow the distribution of the inner disk edge. Adopting the magnetospheric truncation picture \citep{ShapiroTeukolsky1986,Batygin_innerdisk}, the
rotation period distribution of young stars (e.g. in Rho Ophiuchus, $\sim$1-Myr-old which is still in the disk-hosting stage, \citealt{Rebull2018}) should be a good proxy for the orbital period distribution of the innermost planets of a resonant chain system.

For planetary systems whose innermost planets were longer than 20 days, we introduced additional inner planets. We chose a threshold of 20 days because 99\% of young stars in Rho Ophiuchus have rotation period faster than this (Fig. \ref{fig:2，1}). Therefore, 20 days serves as a conservative estimate for the upper limit of the disk's inner edge, which in turn determines the orbital periods of innermost planets in multi-planet systems parked at the disk's inner edge.

The additional planets introduced were assigned the same mass, radius, and envelope fraction as the observed innermost planet. Their orbital periods were drawn randomly, assuming they were in first-order MMR (2:1, 3:2 etc) with longer period planets. We added up to two additional planets if the orbital period of the first planet added is still beyond 20 days. Since tidal effects is stronger for planets with smaller semi-major axes, we calculated the tidal dissipation rate $Q_p^\prime$ on the overlooked planets. We found that an overlooked inner planet could increase log $(Q_p^\prime)$ by $0.9\pm0.3$, i.e. almost one order of magnitude.

We note that the additional planets introduced here could easily evade transit detection if they have a moderate mutual inclination relative to other planets in the system.  For instance, a planet orbiting a Sun-like star on a 10-day orbit would enter a non-transiting configuration with an orbital inclination of 87 deg. A 3-degree mutual inclination is very typical for Kepler-like systems according to previous population-level analysis of transit multiplicity and transit durations \citep{Fabrycky2014,FangMargot2012}. Furthermore, observational evidence indicates that the innermost planet in multi-planet systems often exhibits the largest mutual inclination \citep{Dai2018}. For example, the innermost planet of K2-266 has a mutual inclination $>10$ degree \citep{Rodriguez2018}. The innermost planet Kepler-176 b has a $>3.5$ degree mutual inclination relative to its resonant neighboring planets \citep[see also Fig. \ref{fig:architecture}]{Rowe2014}. Mercury, the innermost planet in the Solar System, is the most inclined \citep[7 degree tilt from the ecliptic][]{Murray}. This phenomenon can be attributed to secular interactions between planets, which redistribute angular momentum deficit \citep{Lithwick2011}. The innermost planet has the lowest angular momentum deficit per unit mass. Given some angular momentum deficit, the innermost planet can be readily excited into an inclined orbit. In the case of the innermost planet in a resonant chain, once it dislodges from mean-motion resonance, its orbital inclination may evolve further due to secular effects or coupling with the stellar quadrupole \citep{Spalding2016}. The enhanced mutual inclination on the innermost planets makes them less likely to be detected by transit surveys such as {\it Kepler} \citep{Borucki+2011} and {\it TESS} \citep{Ricker}.

Once the innermost planet breaks away from MMR, secular interactions dominate over resonant effects and can continue to increase the orbital eccentricity of these planets \citep{Schlaufman2010,Petrovich,Pu}. Consequently, tidal forces may remain active, gradually shrinking the planets' orbits. Eventually, the orbital periods between the innermost planets and longer period companions may may deviate significantly from the initial MMR configuration. This scenario may explain the systems in Fig. \ref{fig:architecture}. Furthermore, once the innermost planets lose the stabilizing influence of MMR, orbital instability may quickly follow. Dynamical simulations by \citet{Li2024} demonstrated that merger events can effectively randomize the period ratios of colliding planets, thereby erasing any trace of the MMR the system may have initially exhibited. Additionally, mutual inclinations generated during close encounters could render the planets non-transiting (or "missing") as discussed earlier in this section.

\subsection{Non-zero Initial \(\Delta\)}

Many disk migration simulations have demonstrated that when planets are captured into first-order MMR during migration, they will end up at a range of initial period ratio deviation $\Delta$ \citep[e.g.][]{Choksi2020,Pichierri,Dai1136}.

 During disk migration, the orbital eccentricity $e$ is damped down by the gaseous disk while it is also pumped up by resonant interaction between the planets. The equilibrium eccentricity is determined by the balance of the $e$ pumping and $e$ damping \citep{TerquemPapaloizou2019}.

 In short, depending on the amount of eccentricity damping in the disk, resonant planets may have a range of initial $e$ and hence a range of non-zero $\Delta$. In a companion paper (Keller \& Dai, in prep), we carried out thousands of convergent disk migrations with a wide range of planetary and disk parameters including different $e$ damping timescales. We found that for planets that engaged into librating first-order MMR during disk migrations, their initial $\Delta$ has a log-normal distribution with a mean at $log_{10} (\Delta)=-3.4\pm0.5$ (Fig. \ref{fig:2，2}).

 On the other hand, the observed near-resonant planets in our sample has a $log_{10} (\Delta)=-1.9\pm0.4$ i.e. the classical 1-2\% deviation from perfect integer ratio \citep{Fabrycky2014}. We compare and contrast the $\Delta$ distribution from disk migration simulations and the observed systems in Fig. \ref{fig:2，2}. \citet{Goldberg2023} proposed that the empirical boundary for separating the librating v.s. circulating planets is at $\Delta=$0.6\% or $log_{10} (\Delta) = -2.2$. Most observed near-resonant planets are in a state of circulation rather than libration.

An inspection of Equation \ref{eq:1} shows that the amount of tidal dissipation \(Q_p^\prime\) is proportional to \((\Delta_f^3 - \Delta_i^3)^{-1}\). In other words, a non-zero $\Delta_i$ would lower the amount of tidal dissipation required to explain the observed $\Delta_f$ distribution. For each near-resonant planet pair in our sample, we bootstrapped 100 times, each time drawing a random \(\Delta_i\) from our disk migration simulations along with other uncertainties such as planet mass and system age. We numerically integrated Equation \ref{eq:1} to determine the change of \(Q_p^\prime\).

The initial and observed $\Delta$ distribution differs by roughly 1.5 orders of magnitude ($log_{10} (\Delta_f)=-3.4\pm0.5$ v.s. $log_{10} (\Delta_f)=-1.9\pm0.4$). Moreover, \(Q_p^\prime\) is proportional to \((\Delta_f^3 - \Delta_i^3)^{-1}\), the third power further accentuates the difference between $\Delta_f$ and $\Delta_i$. Our results suggest that non-zero initial $\Delta_i$ has minimal impact on \(Q_p^\prime\). It can only increase
\(Q_p^\prime\) by no more than a factor of two (Fig. \ref{fig:4}).

\section{Discussion}
\subsection{Resonant repulsion is not the full story}

Our resonant repulsion model successfully reproduced results from the literature \citep{Lithwick_repulsion,LeeMH,Silburt,Millholland_obliquity} namely the traditional treatment of eccentricity tides requires extremely rapid tidal dissipation rates to explain the observed $\Delta$ distribution. Specifically, a reduced tidal quality factor of $log_{10}(Q_p^\prime)=0.1\pm2.1$ is needed for super-Earths and terrestrial-sized planets ($<2R_\oplus$), and $log_{10}(Q_p^\prime)=1.5\pm1.6$ is needed for mini-Neptunes ($>2R_\oplus$). These values are clearly at odds with the Solar System estimates of $log_{10}(Q_p^\prime)=3$ and 5 for the terrestrial planets and icy giant planets respectively \citep{Murray}. See the right panel of Fig. \ref{fig:4}.

The three additional effects investigated in this paper can moderately enhance the effect of resonant repulsion driven by eccentricity tides:

\begin{itemize}
    \item We found that accounting for inflated radii on young planets may increase log $(Q_p^\prime)$ by 0.11$\pm$0.08 for mini-Neptunes ($>2R_\oplus$), or 0.5$\pm$0.2 for super-Earths ($<2R_\oplus$) (Fig. \ref{fig:4}). In other words, radius inflation could increase the required tidal quality factor by half an order of magnitude.
    \item  By including additional planets that lie closer to the host star than the observed resonant pairs, the tidal quality factor \(log_{10}Q_p^\prime\) can increase by $0.9 \pm0.3$ i.e. almost an order of magnitude.
    \item Disk migration deposit planets into MMR with non-zero initial $\Delta$. However, according to our disk migration simulations (Keller \& Dai, in prep), the initial $\Delta_i$ are generally small $log_{10} (\Delta_i)=-3.4\pm0.5$. Including such small initial $\Delta_i$ in resonant repulsion evolution can only increase \(log_{10}Q_p^\prime\) by $<0.1$ in 95\% of the cases. See the left panel of Fig. \ref{fig:4}.

\end{itemize}

Our resonant repulsion model incorporates all three effects simultaneously. With these effects included, the tidal quality factor increases to $log_{10}(Q_p^\prime)=1.2\pm2.1$ for the super-Earths and terrestrial-sized planets, and $log_{10}(Q_p^\prime)=2.4\pm1.7$ for mini-Neptunes ($>2R_\oplus$). This is an improvement over previous estimates in simple eccentricity tidal models. However, our $Q_p^\prime$ are still discrepant from the Solar System values of $log_{10}(Q_p^\prime)=3$ and 5 for rocky and gaseous planets. See the right panel of Fig. \ref{fig:4}.


We further highlight that the near resonant planetary systems HD 109833 \citep[$\sim 20$-Myr-old, a member of the Lower Centaurus Crux, ][]{Wood} and V1298 Tau \citep[$\sim 20$-Myr-old, a member of the Taurus-Auriga, ][]{David2019}. These systems are so young that eccentricity tides likely have not had sufficient time to operate. Yet, the near resonant pairs in  HD109833 and V1298 Tau (Livingston et al. in prep) can have deviation $\Delta =0.8\%-1\%$ from perfect integer ratios. Such significant $\Delta$ at such a young age cannot be explained by resonant repulsion driven by eccentricity tides.

Another limitation of the resonant repulsion model due to eccentricity tides is its inconsistency with observed transit timing variation  (TTV) systems. Specifically, planetary systems with TTVs often exhibit non-zero TTV phases \citep{Lithwick_ttv,Hadden2017,Choksi2023,Goldberg_2023}. TTV phases are a proxy for the free eccentricities of near-resonant planets, with substantial free eccentricities generally producing measurable TTV phases. In order to generate the observed $\Delta$ of $\sim1\%$ in resonant repulsion, a planetary system must evolve over 10s to 100s times $\tau_e$ the eccentricity damping timescales \citep{Lithwick_repulsion,Goldberg2021}. This process effectively eliminates free eccentricities, resulting in negligible TTV phases.  Yet, non-zero TTV phases are commonly seen in TTV systems \citep{Choksi2023,Goldberg_2023}.  This discrepancy further underscores the limitations of resonant repulsion due to eccentricity tides in explaining the near-resonant systems.

Thus, resonant repulsion driven solely by eccentricity tides cannot fully account for the observed orbital architecture of near-resonant planetary systems. This compels us to explore alternative mechanisms for moving planetary systems away from mean-motion resonances. Recently, \citet{Millholland_obliquity} proposed obliquity tides as a more efficient source of tidal dissipation. They argued that resonant planets are more likely to be locked into a high-obliquity Cassini state due to disk migration and the resulting changes in secular frequencies. In a high-obliquity configuration, the tidal bulge continues to shift within the planet's frame, allowing for efficient dissipation of orbital energy even at low orbital eccentricities. As a result, resonant repulsion driven by obliquity tides appears more plausible than that driven by eccentricity tides alone. Notably, the observed orbital architecture of Kepler-221 is most consistent with the rapid evolution under obliquity tides \citep{Goldberg2021}. For a broader population-level analysis of obliquity tides, see \citet{louden2023tidaldissipationregimesshortperiod}.

Turbulence in protoplanetary disks may induce stochastic forcing during Type-I migration \citep{Rein2012,Goldberg2023,Wu_Chen_Lin2024}. This stochastic forcing tends to broaden the distribution of $\Delta$. However, it remains unclear whether turbulent migration models can reproduce the well-known asymmetric distribution of $\Delta$: a trough of planets with $\Delta$ just below 0 and a pile-up of planets with positive 1-2\% deviation \citep{Fabrycky2014}.

During the final stages of protoplanetary disk evolution, the inner disk edge may undergo outward expansion \citep{Liu2017,Hansen2024}. The asymmetric profile at the disk edge generates diverging torques on neighboring planets. \citet{Liu2017} and \citet{Hansen2024} proposed that these torques could pull initially resonant planets out of resonance. Similarly, \citet{Ormel2017} suggested that this mechanism might explain the architecture of the TRAPPIST-1 system, specifically by removing the inner two planets from first-order MMR while maintaining their 3-body Laplace resonance. However, the extent to which this mechanism affects resonant planets located far from the disk’s inner edge remains to be explored.

The cumulative effect of scattering by planetesimals may also account for the observed distributions of $\Delta$ and the non-zero free eccentricity \citep{Chatterjee_2015,Raymond2021,Wu2024}. To produce the observed 1-2\% $\Delta$, the total mass of scattered planetesimals typically needs to be a few percent of the planet's mass.  \citet{Raymond2021} cautioned that resonant-chain systems are so delicate that this amount of planetesimal scattering could destabilize the entire system. \citet{Li2024} argued that orbital instability might play a significant role in shaping the observed orbital architecture of near resonant planets. Contrary to conventional wisdom, \citet{Li2024} showed that merger events do not entirely eliminate near-resonant pairs in a planetary system. Instead, planets that were not involved in mergers tend to accumulate at period ratios slightly wider than resonance.

\citet{Lin2024arXiv} recently proposed that divergent encounters between near-resonant planets tend to push them wider of resonance while generating significant orbital eccentricities. However, orchestrating such encounters for all planets in a resonant chain presents a significant challenge especially without inducing orbital instability.

As discussed above, a variety of alternative theories have been proposed to explain the observed orbital architecture of near-resonant planets. These include orbital instability \citep{Li2024}, disk turbulence \citep{Goldberg2023}, disk edge expansion \citep{Liu2017,Hansen2024}, obliquity tides \citep{Millholland_obliquity,louden2023tidaldissipationregimesshortperiod}, planetesimal scatterings \citep{Chatterjee_2015, Wu2024}, and divergent encounter \citep{Lin2024arXiv}. Our failed attempt to attribute the observed architecture solely to resonant repulsion driven by eccentricity tides suggests that one or more of these alternative mechanisms must play a significant role. More careful investigations are required to constrain the relative contributions of the alternative dynamical processes. 

\subsection{Caveats and Future Work}
We now list some shortcomings of our work and suggest potential avenues for future studies:
\begin{itemize}
    \item Initial condition: In this work we assumed that all planet pairs begin with MMR. However, it remains to be tested whether young planetary systems ($\sim$10-Myr-old) that have recently completed planet formation are still in MMR or have already been dislodged.  If mechanisms such as disk turbulence \citep{Goldberg2023} and disk edge expansion \citep{Liu2017,Hansen2024} are primarily responsible for removing systems from MMR, $\sim$10-Myr-old systems should already be dislodged. Conversely, other mechanisms such as orbital instability \citep{Li2024}, obliquity tides \citep{Millholland_obliquity,louden2023tidaldissipationregimesshortperiod}, and planetesimal scatterings \citep{Chatterjee_2015, Wu2024} may require 10s to 100s Myr to manifest. The discovery and detailed characterization of planetary systems younger than $\sim$10-Myr-old will be instrumental in resolving this question.

    \item Self-consistent radius evolution: Our exploration of radius inflation did not account for the interplay between resonant repulsion and radius evolution. It is plausible that tidal dissipation may sustain an inflated planetary radius for an extended period. Observational evidence seems to support this hypothesis. For example, GJ 436 b \citep{Morley} and WASP-107 b \citep{Sing2024,Welbanks2024,Yu_and_Dai} both have inflated planetary radii that are attributable to eccentricity tides. Furthermore, it has been noted that near-resonant planets are generally puffier than their non-resonant counterparts \citep{Millholland2019,Leleu2024}. Future studies should explore models that couple resonant repulsion with planetary radius evolution.

    \item Direct N-body Simulations: In this work, we modeled resonant repulsion using Eqn. \ref{eq:1} which is strictly valid for isolated pairs of near-resonant planets. A more accurate approach would involve full N-body simulations with eccentricity tides, such as those implemented in {\sc REBOUNDx} \citep{Tamayo_x}. A comprehensive suite of N-body simulations could simultaneously account for dynamical instabilities and other processes which may all play a role in shaping the final orbital architecture.

    \item Higher-order resonances: Our paper exclusively focused on first-order MMRs, as traditional resonant repulsion theory was derived for these resonances \citep{Lithwick_repulsion,Batygin_repulsion}. Higher-order MMRs are not expected to show a positive period ratio deviation $\Delta$ under tidal dissipation \citep{Bailey}. However, the $\Delta$ distribution near higher-order resonance still encodes valuable information about the dynamical evolution of planetary systems and warrants further investigation.
\end{itemize}

\section {Conclusion}

Resonant repulsion coupled with eccentricity tides was among the first explanations proposed for the positive period ratio deviation $\Delta$ observed in planet pairs near first-order MMRs \citep{Papaloizou2010,Lithwick_repulsion,Batygin_repulsion}. However, the required tidal dissipation rate is often unphysically small $Q_p^\prime \lesssim 10$ for some observed near-resonant systems.  In this study, we explored three potential effects that could reduce the required tidal dissipation rate or, equivalently, increase $Q_p^\prime$.

We found that incorporating radius inflation could raise the required tidal quality factor by approximately half an order of magnitude. Including additional, undetected planets closer to the host star could further increase $Q_p^\prime$  by almost an order of magnitude. Finally, the initial period ratio deviation  $\Delta_i$ from disk migration simulations is so small $log_{10} (\Delta_i)=-3.4\pm0.5$ that they have a negligible impact on$Q_p^\prime$.

When combining all three effects, the tidal quality factor increases to $log_{10}(Q_p^\prime)=1.2\pm2.1$ for the super-Earths and terrestrial-sized planets, and $log_{10}(Q_p^\prime)=2.4\pm1.7$ for mini-Neptunes ($>2R_\oplus$). Despite these improvements, these values remain inconsistent with Solar System estimates of $log_{10}(Q_p^\prime)=3$ and 5 for rocky and gaseous planets.

Thus, eccentricity tides alone cannot fully explain the observed $\Delta$ distribution as well as the non-zero TTV phases. Other mechanisms, such as obliquity tides, planetesimal scattering, expanding disk inner edges, disk turbulence, divergent encounters, and dynamical instabilities, are likely essential in shaping planetary systems near first-order MMRs.

\bibliography{main}

\begin{thebibliography}{}
\expandafter\ifx\csname natexlab\endcsname\relax\def\natexlab#1{#1}\fi
\providecommand{\url}[1]{\href{#1}{#1}}

\bibitem[{{Bailey} {et~al.}(2022){Bailey}, {Gilbert}, \& {Fabrycky}}]{Bailey}
{Bailey}, N., {Gilbert}, G., \& {Fabrycky}, D. 2022, \aj, 163, 13

\bibitem[{{Barber} {et~al.}(2024){Barber}, {Thao}, {Mann}, {Vanderburg}, {Mori}, {Livingston}, {Fukui}, {Narita}, {Kraus}, {Tofflemire}, {Newton}, {Winn}, {Jenkins}, {Seager}, {Collins}, \& {Twicken}}]{Barber}
{Barber}, M.~G., {Thao}, P.~C., {Mann}, A.~W., {et~al.} 2024, arXiv e-prints, arXiv:2407.04763

\bibitem[{{Batygin} {et~al.}(2023){Batygin}, {Adams}, \& {Becker}}]{Batygin_innerdisk}
{Batygin}, K., {Adams}, F.~C., \& {Becker}, J. 2023, \apjl, 951, L19

\bibitem[{{Batygin} \& {Morbidelli}(2013)}]{Batygin_repulsion}
{Batygin}, K., \& {Morbidelli}, A. 2013, \aj, 145, 1

\bibitem[{{Berger} {et~al.}(2020){Berger}, {Huber}, {Gaidos}, {van Saders}, \& {Weiss}}]{Berger_age}
{Berger}, T.~A., {Huber}, D., {Gaidos}, E., {van Saders}, J.~L., \& {Weiss}, L.~M. 2020, \aj, 160, 108

\bibitem[{{Borucki} {et~al.}(2011){Borucki}, {Koch}, {Basri}, {Batalha}, {Boss}, {Brown}, {Caldwell}, {Christensen-Dalsgaard}, {Cochran}, {DeVore}, {Dunham}, {Dupree}, {Gautier}, {Geary}, {Gilliland}, {Gould}, {Howell}, {Jenkins}, {Kjeldsen}, {Latham}, {Lissauer}, {Marcy}, {Monet}, {Sasselov}, {Tarter}, {Charbonneau}, {Doyle}, {Ford}, {Fortney}, {Holman}, {Seager}, {Steffen}, {Welsh}, {Allen}, {Bryson}, {Buchhave}, {Chandrasekaran}, {Christiansen}, {Ciardi}, {Clarke}, {Dotson}, {Endl}, {Fischer}, {Fressin}, {Haas}, {Horch}, {Howard}, {Isaacson}, {Kolodziejczak}, {Li}, {MacQueen}, {Meibom}, {Prsa}, {Quintana}, {Rowe}, {Sherry}, {Tenenbaum}, {Torres}, {Twicken}, {Van Cleve}, {Walkowicz}, \& {Wu}}]{Borucki+2011}
{Borucki}, W.~J., {Koch}, D.~G., {Basri}, G., {et~al.} 2011, \apj, 728, 117

\bibitem[{{Brasser} {et~al.}(2022){Brasser}, {Pichierri}, {Dobos}, \& {Barr}}]{Brasser2021}
{Brasser}, R., {Pichierri}, G., {Dobos}, V., \& {Barr}, A.~C. 2022, \mnras, 515, 2373

\bibitem[{Chatterjee \& Ford(2015)}]{Chatterjee_2015}
Chatterjee, S., \& Ford, E.~B. 2015, The Astrophysical Journal, 803, 33.
\newblock \url{https://doi.org/10.1088/0004-637x/803/1/33}

\bibitem[{Chen \& Rogers(2016)}]{Chen_2016}
Chen, H., \& Rogers, L.~A. 2016, The Astrophysical Journal, 831, 180.
\newblock \url{http://dx.doi.org/10.3847/0004-637X/831/2/180}

\bibitem[{{Choksi} \& {Chiang}(2020)}]{Choksi2020}
{Choksi}, N., \& {Chiang}, E. 2020, \mnras, 495, 4192

\bibitem[{{Choksi} \& {Chiang}(2023)}]{Choksi2023}
---. 2023, \mnras, 522, 1914

\bibitem[{{Crossfield} {et~al.}(2022){Crossfield}, {Malik}, {Hill}, {Kane}, {Foley}, {Polanski}, {Coria}, {Brande}, {Zhang}, {Wienke}, {Kreidberg}, {Cowan}, {Dragomir}, {Gorjian}, {Mikal-Evans}, {Benneke}, {Christiansen}, {Deming}, \& {Morales}}]{Crossfield2022}
{Crossfield}, I. J.~M., {Malik}, M., {Hill}, M.~L., {et~al.} 2022, \apjl, 937, L17

\bibitem[{{Dai} {et~al.}(2018){Dai}, {Masuda}, \& {Winn}}]{Dai2018}
{Dai}, F., {Masuda}, K., \& {Winn}, J.~N. 2018, \apjl, 864, L38

\bibitem[{{Dai} {et~al.}(2019){Dai}, {Masuda}, {Winn}, \& {Zeng}}]{Dai2019}
{Dai}, F., {Masuda}, K., {Winn}, J.~N., \& {Zeng}, L. 2019, \apj, 883, 79

\bibitem[{{Dai} {et~al.}(2023){Dai}, {Masuda}, {Beard}, {Robertson}, {Goldberg}, {Batygin}, {Bouma}, {Lissauer}, {Knudstrup}, {Albrecht}, {Howard}, {Knutson}, {Petigura}, {Weiss}, {Isaacson}, {Kristiansen}, {Osborn}, {Wang}, {Wang}, {Behmard}, {Greklek-McKeon}, {Vissapragada}, {Batalha}, {Brinkman}, {Chontos}, {Crossfield}, {Dressing}, {Fetherolf}, {Fulton}, {Hill}, {Huber}, {Kane}, {Lubin}, {MacDougall}, {Mayo}, {Mo{\v{c}}nik}, {Akana Murphy}, {Rubenzahl}, {Scarsdale}, {Tyler}, {Zandt}, {Polanski}, {Schwengeler}, {Terentev}, {Benni}, {Bieryla}, {Ciardi}, {Falk}, {Furlan}, {Girardin}, {Guerra}, {Hesse}, {Howell}, {Lillo-Box}, {Matthews}, {Twicken}, {Villase{\~n}or}, {Latham}, {Jenkins}, {Ricker}, {Seager}, {Vanderspek}, \& {Winn}}]{Dai1136}
{Dai}, F., {Masuda}, K., {Beard}, C., {et~al.} 2023, \aj, 165, 33

\bibitem[{Dai {et~al.}(2024)Dai, Goldberg, Batygin, van Saders, Chiang, Choksi, Li, Petigura, Gilbert, Millholland, Dai, Bouma, Weiss, \& Winn}]{dai2024prevalenceresonanceyoungclosein}
Dai, F., Goldberg, M., Batygin, K., {et~al.} 2024, The Prevalence of Resonance Among Young, Close-in Planets, , , arXiv:2406.06885.
\newblock \url{https://arxiv.org/abs/2406.06885}

\bibitem[{{David} {et~al.}(2019){David}, {Petigura}, {Luger}, {Foreman-Mackey}, {Livingston}, {Mamajek}, \& {Hillenbrand}}]{David2019}
{David}, T.~J., {Petigura}, E.~A., {Luger}, R., {et~al.} 2019, \apjl, 885, L12

\bibitem[{{Delisle} \& {Laskar}(2014)}]{Delisle2014}
{Delisle}, J.~B., \& {Laskar}, J. 2014, \aap, 570, L7

\bibitem[{{Dressing} {et~al.}(2015){Dressing}, {Charbonneau}, {Dumusque}, {Gettel}, {Pepe}, {Collier Cameron}, {Latham}, {Molinari}, {Udry}, {Affer}, {Bonomo}, {Buchhave}, {Cosentino}, {Figueira}, {Fiorenzano}, {Harutyunyan}, {Haywood}, {Johnson}, {Lopez-Morales}, {Lovis}, {Malavolta}, {Mayor}, {Micela}, {Motalebi}, {Nascimbeni}, {Phillips}, {Piotto}, {Pollacco}, {Queloz}, {Rice}, {Sasselov}, {S{\'e}gransan}, {Sozzetti}, {Szentgyorgyi}, \& {Watson}}]{Dressing}
{Dressing}, C.~D., {Charbonneau}, D., {Dumusque}, X., {et~al.} 2015, \apj, 800, 135

\bibitem[{{Emsenhuber} {et~al.}(2021){Emsenhuber}, {Mordasini}, {Burn}, {Alibert}, {Benz}, \& {Asphaug}}]{Emsenhuber2021}
{Emsenhuber}, A., {Mordasini}, C., {Burn}, R., {et~al.} 2021, \aap, 656, A69

\bibitem[{{Fabrycky} {et~al.}(2014){Fabrycky}, {Lissauer}, {Ragozzine}, {Rowe}, {Steffen}, {Agol}, {Barclay}, {Batalha}, {Borucki}, {Ciardi}, {Ford}, {Gautier}, {Geary}, {Holman}, {Jenkins}, {Li}, {Morehead}, {Morris}, {Shporer}, {Smith}, {Still}, \& {Van Cleve}}]{Fabrycky2014}
{Fabrycky}, D.~C., {Lissauer}, J.~J., {Ragozzine}, D., {et~al.} 2014, \apj, 790, 146

\bibitem[{{Fang} \& {Margot}(2012)}]{FangMargot2012}
{Fang}, J., \& {Margot}, J.-L. 2012, \apj, 761, 92

\bibitem[{{Ford} {et~al.}(2012){Ford}, {Fabrycky}, {Steffen}, {Carter}, {Fressin}, {Holman}, {Lissauer}, {Moorhead}, {Morehead}, {Ragozzine}, {Rowe}, {Welsh}, {Allen}, {Batalha}, {Borucki}, {Bryson}, {Buchhave}, {Burke}, {Caldwell}, {Charbonneau}, {Clarke}, {Cochran}, {D{\'e}sert}, {Endl}, {Everett}, {Fischer}, {Gautier}, {Gilliland}, {Jenkins}, {Haas}, {Horch}, {Howell}, {Ibrahim}, {Isaacson}, {Koch}, {Latham}, {Li}, {Lucas}, {MacQueen}, {Marcy}, {McCauliff}, {Mullally}, {Quinn}, {Quintana}, {Shporer}, {Still}, {Tenenbaum}, {Thompson}, {Torres}, {Twicken}, {Wohler}, \& {Kepler Science Team}}]{Ford2012}
{Ford}, E.~B., {Fabrycky}, D.~C., {Steffen}, J.~H., {et~al.} 2012, \apj, 750, 113

\bibitem[{{Fortney} {et~al.}(2007){Fortney}, {Marley}, \& {Barnes}}]{Fortney}
{Fortney}, J.~J., {Marley}, M.~S., \& {Barnes}, J.~W. 2007, \apj, 659, 1661

\bibitem[{{Fulton} {et~al.}(2017){Fulton}, {Petigura}, {Howard}, {Isaacson}, {Marcy}, {Cargile}, {Hebb}, {Weiss}, {Johnson}, {Morton}, {Sinukoff}, {Crossfield}, \& {Hirsch}}]{Fulton}
{Fulton}, B.~J., {Petigura}, E.~A., {Howard}, A.~W., {et~al.} 2017, \aj, 154, 109

\bibitem[{{Ginzburg} {et~al.}(2018){Ginzburg}, {Schlichting}, \& {Sari}}]{Ginzburg}
{Ginzburg}, S., {Schlichting}, H.~E., \& {Sari}, R. 2018, \mnras, 476, 759

\bibitem[{Goldberg \& Batygin(2021)}]{Goldberg2021}
Goldberg, M., \& Batygin, K. 2021, doi:10.3847/1538-3881/abfb78.
\newblock \url{http://arxiv.org/abs/2105.07368 http://dx.doi.org/10.3847/1538-3881/abfb78}

\bibitem[{{Goldberg} \& {Batygin}(2023)}]{Goldberg2023}
{Goldberg}, M., \& {Batygin}, K. 2023, \apj, 948, 12

\bibitem[{Goldberg \& Batygin(2023)}]{Goldberg_2023}
Goldberg, M., \& Batygin, K. 2023, The Astrophysical Journal, 948, 12.
\newblock \url{http://dx.doi.org/10.3847/1538-4357/acc9ae}

\bibitem[{{Goldberg} {et~al.}(2022){Goldberg}, {Batygin}, \& {Morbidelli}}]{Goldberg_stability}
{Goldberg}, M., {Batygin}, K., \& {Morbidelli}, A. 2022, arXiv e-prints, arXiv:2207.13833

\bibitem[{{Goldreich} \& {Soter}(1966)}]{Goldreich_Soter1966}
{Goldreich}, P., \& {Soter}, S. 1966, \icarus, 5, 375

\bibitem[{{Goldreich} \& {Tremaine}(1979)}]{Goldreich1979}
{Goldreich}, P., \& {Tremaine}, S. 1979, \apj, 233, 857

\bibitem[{{Hadden} \& {Lithwick}(2014)}]{HaddenLithwick2014}
{Hadden}, S., \& {Lithwick}, Y. 2014, \apj, 787, 80

\bibitem[{{Hadden} \& {Lithwick}(2017)}]{Hadden2017}
---. 2017, \aj, 154, 5

\bibitem[{{Hamer} \& {Schlaufman}(2024)}]{Hamer2024}
{Hamer}, J.~H., \& {Schlaufman}, K.~C. 2024, \aj, 167, 55

\bibitem[{{Hansen} {et~al.}(2024){Hansen}, {Yu}, \& {Hasegawa}}]{Hansen2024}
{Hansen}, B. M.~S., {Yu}, T.-Y., \& {Hasegawa}, Y. 2024, arXiv e-prints, arXiv:2405.12388

\bibitem[{{Hu} {et~al.}(2024){Hu}, {Bello-Arufe}, {Zhang}, {Paragas}, {Zilinskas}, {van Buchem}, {Bess}, {Patel}, {Ito}, {Damiano}, {Scheucher}, {Oza}, {Knutson}, {Miguel}, {Dragomir}, {Brandeker}, \& {Demory}}]{Hu2024}
{Hu}, R., {Bello-Arufe}, A., {Zhang}, M., {et~al.} 2024, arXiv e-prints, arXiv:2405.04744

\bibitem[{{Huang} \& {Ormel}(2022)}]{Huang_Ormel}
{Huang}, S., \& {Ormel}, C.~W. 2022, \mnras, 511, 3814

\bibitem[{{Izidoro} {et~al.}(2017){Izidoro}, {Ogihara}, {Raymond}, {Morbidelli}, {Pierens}, {Bitsch}, {Cossou}, \& {Hersant}}]{Izidoro}
{Izidoro}, A., {Ogihara}, M., {Raymond}, S.~N., {et~al.} 2017, \mnras, 470, 1750

\bibitem[{{Kreidberg} {et~al.}(2019){Kreidberg}, {Koll}, {Morley}, {Hu}, {Schaefer}, {Deming}, {Stevenson}, {Dittmann}, {Vanderburg}, {Berardo}, {Guo}, {Stassun}, {Crossfield}, {Charbonneau}, {Latham}, {Loeb}, {Ricker}, {Seager}, \& {Vanderspek}}]{Kreidberg2019}
{Kreidberg}, L., {Koll}, D. D.~B., {Morley}, C., {et~al.} 2019, \nat, 573, 87

\bibitem[{{Lee}(2019)}]{Lee2019}
{Lee}, E.~J. 2019, \apj, 878, 36

\bibitem[{{Lee} {et~al.}(2013){Lee}, {Fabrycky}, \& {Lin}}]{LeeMH}
{Lee}, M.~H., {Fabrycky}, D., \& {Lin}, D.~N.~C. 2013, \apj, 774, 52

\bibitem[{{Leleu} {et~al.}(2024){Leleu}, {Delisle}, {Burn}, {Izidoro}, {Udry}, {Dumusque}, {Lovis}, {Millholland}, {Parc}, {Bouchy}, {Bourrier}, {Alibert}, {Faria}, {Mordasini}, \& {S{\'e}gransan}}]{Leleu2024}
{Leleu}, A., {Delisle}, J.-B., {Burn}, R., {et~al.} 2024, \aap, 687, L1

\bibitem[{{Li} {et~al.}(2024){Li}, {Chiang}, {Choksi}, \& {Dai}}]{Li2024}
{Li}, R., {Chiang}, E., {Choksi}, N., \& {Dai}, F. 2024, arXiv e-prints, arXiv:2408.10206

\bibitem[{{Lin} \& {Papaloizou}(1986)}]{Lin1986}
{Lin}, D.~N.~C., \& {Papaloizou}, J. 1986, \apj, 309, 846

\bibitem[{{Lin} {et~al.}(2024){Lin}, {Dudiak}, {Hadden}, \& {Tamayo}}]{Lin2024arXiv}
{Lin}, J., {Dudiak}, I., {Hadden}, S., \& {Tamayo}, D. 2024, arXiv e-prints, arXiv:2412.12415

\bibitem[{{Lithwick} \& {Wu}(2011)}]{Lithwick2011}
{Lithwick}, Y., \& {Wu}, Y. 2011, \apj, 739, 31

\bibitem[{Lithwick \& Wu(2012)}]{Lithwick_repulsion}
Lithwick, Y., \& Wu, Y. 2012, The Astrophysical Journal, 756, L11.
\newblock \url{https://iopscience.iop.org/article/10.1088/2041-8205/756/1/L11}

\bibitem[{{Lithwick} {et~al.}(2012){Lithwick}, {Xie}, \& {Wu}}]{Lithwick_ttv}
{Lithwick}, Y., {Xie}, J., \& {Wu}, Y. 2012, \apj, 761, 122

\bibitem[{{Liu} {et~al.}(2017){Liu}, {Ormel}, \& {Lin}}]{Liu2017}
{Liu}, B., {Ormel}, C.~W., \& {Lin}, D. N.~C. 2017, \aap, 601, A15

\bibitem[{{Livingston} {et~al.}(2018){Livingston}, {Crossfield}, {Petigura}, {Gonzales}, {Ciardi}, {Beichman}, {Christiansen}, {Dressing}, {Henning}, {Howard}, {Isaacson}, {Fulton}, {Kosiarek}, {Schlieder}, {Sinukoff}, \& {Tamura}}]{Livingston2018}
{Livingston}, J.~H., {Crossfield}, I. J.~M., {Petigura}, E.~A., {et~al.} 2018, \aj, 156, 277

\bibitem[{Louden {et~al.}(2023)Louden, Laughlin, \& Millholland}]{louden2023tidaldissipationregimesshortperiod}
Louden, E., Laughlin, G., \& Millholland, S. 2023, Tidal Dissipation Regimes Among the Short-Period Exoplanets, , , arXiv:2311.03576.
\newblock \url{https://arxiv.org/abs/2311.03576}

\bibitem[{{Luque} {et~al.}(2023){Luque}, {Osborn}, {Leleu}, {Pall{\'e}}, {Bonfanti}, {Barrag{\'a}n}, {Wilson}, {Broeg}, {Cameron}, {Lendl}, {Maxted}, {Alibert}, {Gandolfi}, {Delisle}, {Hooton}, {Egger}, {Nowak}, {Lafarga}, {Rapetti}, {Twicken}, {Morales}, {Carleo}, {Orell-Miquel}, {Adibekyan}, {Alonso}, {Alqasim}, {Amado}, {Anderson}, {Anglada-Escud{\'e}}, {Bandy}, {B{\'a}rczy}, {Barrado Navascues}, {Barros}, {Baumjohann}, {Bayliss}, {Bean}, {Beck}, {Beck}, {Benz}, {Billot}, {Bonfils}, {Borsato}, {Boyle}, {Brandeker}, {Bryant}, {Cabrera}, {Carrazco-Gaxiola}, {Charbonneau}, {Charnoz}, {Ciardi}, {Cochran}, {Collins}, {Crossfield}, {Csizmadia}, {Cubillos}, {Dai}, {Davies}, {Deeg}, {Deleuil}, {Deline}, {Delrez}, {Demangeon}, {Demory}, {Ehrenreich}, {Erikson}, {Esparza-Borges}, {Falk}, {Fortier}, {Fossati}, {Fridlund}, {Fukui}, {Garcia-Mejia}, {Gill}, {Gillon}, {Goffo}, {G{\'o}mez Maqueo Chew}, {G{\"u}del}, {Guenther}, {G{\"u}nther}, {Hatzes}, {Helling}, {Hesse}, {Howell}, {Hoyer}, {Ikuta}, {Isaak}, {Jenkins},
  {Kagetani}, {Kiss}, {Kodama}, {Korth}, {Lam}, {Laskar}, {Latham}, {Lecavelier des Etangs}, {Leon}, {Livingston}, {Magrin}, {Matson}, {Matthews}, {Mordasini}, {Mori}, {Moyano}, {Munari}, {Murgas}, {Narita}, {Nascimbeni}, {Olofsson}, {Osborne}, {Ottensamer}, {Pagano}, {Parviainen}, {Peter}, {Piotto}, {Pollacco}, {Queloz}, {Quinn}, {Quirrenbach}, {Ragazzoni}, {Rando}, {Ratti}, {Rauer}, {Redfield}, {Ribas}, {Ricker}, {Rudat}, {Sabin}, {Salmon}, {Santos}, {Scandariato}, {Schanche}, {Schlieder}, {Seager}, {S{\'e}gransan}, {Shporer}, {Simon}, {Smith}, {Sousa}, {Stalport}, {Szab{\'o}}, {Thomas}, {Tuson}, {Udry}, {Vanderburg}, {Van Eylen}, {Van Grootel}, {Venturini}, {Walter}, {Walton}, {Watanabe}, {Winn}, \& {Zingales}}]{Luque}
{Luque}, R., {Osborn}, H.~P., {Leleu}, A., {et~al.} 2023, \nat, 623, 932

\bibitem[{{MacDonald} {et~al.}(2016){MacDonald}, {Ragozzine}, {Fabrycky}, {Ford}, {Holman}, {Isaacson}, {Lissauer}, {Lopez}, {Mazeh}, {Rogers}, {Rowe}, {Steffen}, \& {Torres}}]{MacDonald2016}
{MacDonald}, M.~G., {Ragozzine}, D., {Fabrycky}, D.~C., {et~al.} 2016, \aj, 152, 105

\bibitem[{{Mann} {et~al.}(2022){Mann}, {Wood}, {Schmidt}, {Barber}, {Owen}, {Tofflemire}, {Newton}, {Mamajek}, {Bush}, {Mace}, {Kraus}, {Thao}, {Vanderburg}, {Llama}, {Johns-Krull}, {Prato}, {Stahl}, {Tang}, {Fields}, {Collins}, {Collins}, {Gan}, {Jensen}, {Kamler}, {Schwarz}, {Furlan}, {Gnilka}, {Howell}, {Lester}, {Owens}, {Suarez}, {Mekarnia}, {Guillot}, {Abe}, {Triaud}, {Johnson}, {Milburn}, {Rizzuto}, {Quinn}, {Kerr}, {Ricker}, {Vanderspek}, {Latham}, {Seager}, {Winn}, {Jenkins}, {Guerrero}, {Shporer}, {Schlieder}, {McLean}, \& {Wohler}}]{Mann1227}
{Mann}, A.~W., {Wood}, M.~L., {Schmidt}, S.~P., {et~al.} 2022, \aj, 163, 156

\bibitem[{Masset {et~al.}(2006)Masset, Morbidelli, Crida, \& Ferreira}]{Masset_2006}
Masset, F.~S., Morbidelli, A., Crida, A., \& Ferreira, J. 2006, The Astrophysical Journal, 642, 478.
\newblock \url{https://doi.org/10.1086/500967}

\bibitem[{{Matsumoto} \& {Ogihara}(2020)}]{Matsumoto}
{Matsumoto}, Y., \& {Ogihara}, M. 2020, \apj, 893, 43

\bibitem[{{McNeil} {et~al.}(2005){McNeil}, {Duncan}, \& {Levison}}]{McNeil}
{McNeil}, D., {Duncan}, M., \& {Levison}, H.~F. 2005, \aj, 130, 2884

\bibitem[{{Millholland}(2019)}]{Millholland2019}
{Millholland}, S. 2019, \apj, 886, 72

\bibitem[{{Millholland} \& {Laughlin}(2019)}]{Millholland_obliquity}
{Millholland}, S., \& {Laughlin}, G. 2019, Nature Astronomy, 3, 424

\bibitem[{{Mills} {et~al.}(2016){Mills}, {Fabrycky}, {Migaszewski}, {Ford}, {Petigura}, \& {Isaacson}}]{MillsNature}
{Mills}, S.~M., {Fabrycky}, D.~C., {Migaszewski}, C., {et~al.} 2016, \nat, 533, 509

\bibitem[{{Morley} {et~al.}(2017){Morley}, {Knutson}, {Line}, {Fortney}, {Thorngren}, {Marley}, {Teal}, \& {Lupu}}]{Morley}
{Morley}, C.~V., {Knutson}, H., {Line}, M., {et~al.} 2017, \aj, 153, 86

\bibitem[{{Murray} \& {Dermott}(1999)}]{Murray}
{Murray}, C.~D., \& {Dermott}, S.~F. 1999, {Solar system dynamics}

\bibitem[{{NASA Exoplanet Science Institute}(2020)}]{https://doi.org/10.26133/nea12}
{NASA Exoplanet Science Institute}. 2020, Planetary Systems Table,  IPAC, doi:10.26133/NEA12.
\newblock \url{https://catcopy.ipac.caltech.edu/dois/doi.php?id=10.26133/NEA12}

\bibitem[{{Nelson}(2018)}]{Nelson2018}
{Nelson}, R.~P. 2018, in Handbook of Exoplanets, ed. H.~J. {Deeg} \& J.~A. {Belmonte}, 139

\bibitem[{{Ogihara} {et~al.}(2018){Ogihara}, {Kokubo}, {Suzuki}, \& {Morbidelli}}]{Ogihara2018}
{Ogihara}, M., {Kokubo}, E., {Suzuki}, T.~K., \& {Morbidelli}, A. 2018, \aap, 615, A63

\bibitem[{{Ormel} {et~al.}(2017){Ormel}, {Liu}, \& {Schoonenberg}}]{Ormel2017}
{Ormel}, C.~W., {Liu}, B., \& {Schoonenberg}, D. 2017, \aap, 604, A1

\bibitem[{{Otegi} {et~al.}(2020){Otegi}, {Bouchy}, \& {Helled}}]{Otegi_2020}
{Otegi}, J.~F., {Bouchy}, F., \& {Helled}, R. 2020, \aap, 634, A43

\bibitem[{{Owen} \& {Wu}(2017)}]{OwenWu}
{Owen}, J.~E., \& {Wu}, Y. 2017, \apj, 847, 29

\bibitem[{Papaloizou {et~al.}(2017)Papaloizou, Szuszkiewicz, \& Terquem}]{Papaloizou2017}
Papaloizou, J. C.~B., Szuszkiewicz, E., \& Terquem, C. 2017, doi:10.1093/mnras/stx2980.
\newblock \url{http://arxiv.org/abs/1711.07932 http://dx.doi.org/10.1093/mnras/stx2980}

\bibitem[{{Papaloizou} \& {Terquem}(2010)}]{Papaloizou2010}
{Papaloizou}, J. C.~B., \& {Terquem}, C. 2010, \mnras, 405, 573

\bibitem[{{Paxton} {et~al.}(2015){Paxton}, {Marchant}, {Schwab}, {Bauer}, {Bildsten}, {Cantiello}, {Dessart}, {Farmer}, {Hu}, {Langer}, {Townsend}, {Townsley}, \& {Timmes}}]{Paxton2015}
{Paxton}, B., {Marchant}, P., {Schwab}, J., {et~al.} 2015, \apjs, 220, 15

\bibitem[{{Peale}(1976)}]{Peale1976}
{Peale}, S.~J. 1976, \araa, 14, 215

\bibitem[{{Petrovich} {et~al.}(2018){Petrovich}, {Deibert}, \& {Wu}}]{Petrovich}
{Petrovich}, C., {Deibert}, E., \& {Wu}, Y. 2018, ArXiv e-prints, arXiv:1804.05065

\bibitem[{{Pichierri} \& {Morbidelli}(2020)}]{Pichierri2020}
{Pichierri}, G., \& {Morbidelli}, A. 2020, \mnras, 494, 4950

\bibitem[{{Pichierri} {et~al.}(2018){Pichierri}, {Morbidelli}, \& {Crida}}]{Pichierri}
{Pichierri}, G., {Morbidelli}, A., \& {Crida}, A. 2018, Celestial Mechanics and Dynamical Astronomy, 130, 54

\bibitem[{{Plavchan} {et~al.}(2020){Plavchan}, {Barclay}, {Gagn{\'e}}, {Gao}, {Cale}, {Matzko}, {Dragomir}, {Quinn}, {Feliz}, {Stassun}, {Crossfield}, {Berardo}, {Latham}, {Tieu}, {Anglada-Escud{\'e}}, {Ricker}, {Vanderspek}, {Seager}, {Winn}, {Jenkins}, {Rinehart}, {Krishnamurthy}, {Dynes}, {Doty}, {Adams}, {Afanasev}, {Beichman}, {Bottom}, {Bowler}, {Brinkworth}, {Brown}, {Cancino}, {Ciardi}, {Clampin}, {Clark}, {Collins}, {Davison}, {Foreman-Mackey}, {Furlan}, {Gaidos}, {Geneser}, {Giddens}, {Gilbert}, {Hall}, {Hellier}, {Henry}, {Horner}, {Howard}, {Huang}, {Huber}, {Kane}, {Kenworthy}, {Kielkopf}, {Kipping}, {Klenke}, {Kruse}, {Latouf}, {Lowrance}, {Mennesson}, {Mengel}, {Mills}, {Morton}, {Narita}, {Newton}, {Nishimoto}, {Okumura}, {Palle}, {Pepper}, {Quintana}, {Roberge}, {Roccatagliata}, {Schlieder}, {Tanner}, {Teske}, {Tinney}, {Vanderburg}, {von Braun}, {Walp}, {Wang}, {Wang}, {Weigand}, {White}, {Wittenmyer}, {Wright}, {Youngblood}, {Zhang}, \& {Zilberman}}]{Plavchan}
{Plavchan}, P., {Barclay}, T., {Gagn{\'e}}, J., {et~al.} 2020, \nat, 582, 497

\bibitem[{{Pu} \& {Lai}(2019)}]{Pu}
{Pu}, B., \& {Lai}, D. 2019, \mnras, 488, 3568

\bibitem[{Raymond {et~al.}(2021)Raymond, Izidoro, Bolmont, Dorn, Selsis, Turbet, Agol, Barth, Carone, Dasgupta, Gillon, \& Grimm}]{Raymond2021}
Raymond, S.~N., Izidoro, A., Bolmont, E., {et~al.} 2021, doi:10.1038/s41550-021-01518-6.
\newblock \url{http://arxiv.org/abs/2111.13351 http://dx.doi.org/10.1038/s41550-021-01518-6}

\bibitem[{{Rebull} {et~al.}(2018){Rebull}, {Stauffer}, {Cody}, {Hillenbrand}, {David}, \& {Pinsonneault}}]{Rebull2018}
{Rebull}, L.~M., {Stauffer}, J.~R., {Cody}, A.~M., {et~al.} 2018, \aj, 155, 196

\bibitem[{{Rein}(2012)}]{Rein2012}
{Rein}, H. 2012, \mnras, 427, L21

\bibitem[{{Ricker} {et~al.}(2014){Ricker}, {Winn}, {Vanderspek}, {Latham}, {Bakos}, {Bean}, {Berta-Thompson}, {Brown}, {Buchhave}, {Butler}, {Butler}, {Chaplin}, {Charbonneau}, {Christensen-Dalsgaard}, {Clampin}, {Deming}, {Doty}, {De Lee}, {Dressing}, {Dunham}, {Endl}, {Fressin}, {Ge}, {Henning}, {Holman}, {Howard}, {Ida}, {Jenkins}, {Jernigan}, {Johnson}, {Kaltenegger}, {Kawai}, {Kjeldsen}, {Laughlin}, {Levine}, {Lin}, {Lissauer}, {MacQueen}, {Marcy}, {McCullough}, {Morton}, {Narita}, {Paegert}, {Palle}, {Pepe}, {Pepper}, {Quirrenbach}, {Rinehart}, {Sasselov}, {Sato}, {Seager}, {Sozzetti}, {Stassun}, {Sullivan}, {Szentgyorgyi}, {Torres}, {Udry}, \& {Villasenor}}]{Ricker}
{Ricker}, G.~R., {Winn}, J.~N., {Vanderspek}, R., {et~al.} 2014, Society of Photo-Optical Instrumentation Engineers (SPIE) Conference Series, Vol. 9143, {Transiting Exoplanet Survey Satellite (TESS)}, 914320

\bibitem[{{Rodriguez} {et~al.}(2018){Rodriguez}, {Becker}, {Eastman}, {Hadden}, {Vanderburg}, {Khain}, {Quinn}, {Mayo}, {Dressing}, {Schlieder}, {Ciardi}, {Latham}, {Rappaport}, {Adams}, {Berlind}, {Bieryla}, {Calkins}, {Esquerdo}, {Kristiansen}, {Omohundro}, {Schwengeler}, {Stassun}, \& {Terentev}}]{Rodriguez2018}
{Rodriguez}, J.~E., {Becker}, J.~C., {Eastman}, J.~D., {et~al.} 2018, ArXiv e-prints, arXiv:1806.08368

\bibitem[{{Rogers}(2015)}]{Rogers}
{Rogers}, L.~A. 2015, \apj, 801, 41

\bibitem[{{Rowe} {et~al.}(2014){Rowe}, {Bryson}, {Marcy}, {Lissauer}, {Jontof-Hutter}, {Mullally}, {Gilliland}, {Issacson}, {Ford}, {Howell}, {Borucki}, {Haas}, {Huber}, {Steffen}, {Thompson}, {Quintana}, {Barclay}, {Still}, {Fortney}, {Gautier}, {Hunter}, {Caldwell}, {Ciardi}, {Devore}, {Cochran}, {Jenkins}, {Agol}, {Carter}, \& {Geary}}]{Rowe2014}
{Rowe}, J.~F., {Bryson}, S.~T., {Marcy}, G.~W., {et~al.} 2014, \apj, 784, 45

\bibitem[{{Schlaufman}(2010)}]{Schlaufman2010}
{Schlaufman}, K.~C. 2010, \apj, 719, 602

\bibitem[{{Shapiro} \& {Teukolsky}(1986)}]{ShapiroTeukolsky1986}
{Shapiro}, S.~L., \& {Teukolsky}, S.~A. 1986, {Black Holes, White Dwarfs and Neutron Stars: The Physics of Compact Objects}

\bibitem[{Silburt \& Rein(2015)}]{Silburt}
Silburt, A., \& Rein, H. 2015, Monthly Notices of the Royal Astronomical Society, 453, 4089.
\newblock \url{https://doi.org/10.1093/mnras/stv1924}

\bibitem[{{Sing} {et~al.}(2024){Sing}, {Rustamkulov}, {Thorngren}, {Barstow}, {Tremblin}, {Alves de Oliveira}, {Beck}, {Birkmann}, {Challener}, {Crouzet}, {Espinoza}, {Ferruit}, {Giardino}, {Gressier}, {Lee}, {Lewis}, {Maiolino}, {Manjavacas}, {Rauscher}, {Sirianni}, \& {Valenti}}]{Sing2024}
{Sing}, D.~K., {Rustamkulov}, Z., {Thorngren}, D.~P., {et~al.} 2024, \nat, 630, 831

\bibitem[{{Spalding} \& {Batygin}(2016)}]{Spalding2016}
{Spalding}, C., \& {Batygin}, K. 2016, \apj, 830, 5

\bibitem[{{Tamayo} {et~al.}(2020){Tamayo}, {Rein}, {Shi}, \& {Hernandez}}]{Tamayo_x}
{Tamayo}, D., {Rein}, H., {Shi}, P., \& {Hernandez}, D.~M. 2020, \mnras, 491, 2885

\bibitem[{Terquem \& Papaloizou(2007)}]{Terquem_2007}
Terquem, C., \& Papaloizou, J. C.~B. 2007, The Astrophysical Journal, 654, 1110.
\newblock \url{https://doi.org/10.1086/509497}

\bibitem[{{Terquem} \& {Papaloizou}(2019)}]{TerquemPapaloizou2019}
{Terquem}, C., \& {Papaloizou}, J. C.~B. 2019, \mnras, 482, 530

\bibitem[{{Tittemore} \& {Wisdom}(1990)}]{Tittemore}
{Tittemore}, W.~C., \& {Wisdom}, J. 1990, \icarus, 85, 394

\bibitem[{{Vach} {et~al.}(2024){Vach}, {Zhou}, {Huang}, {Rogers}, {Bouma}, {Douglas}, {Kunimoto}, {Mann}, {Barber}, {Quinn}, {Latham}, {Bieryla}, \& {Collins}}]{Vach}
{Vach}, S., {Zhou}, G., {Huang}, C.~X., {et~al.} 2024, \aj, 167, 210

\bibitem[{{Wang} \& {Lin}(2023)}]{Wang2023}
{Wang}, S., \& {Lin}, D.~N.~C. 2023, \aj, 165, 174

\bibitem[{Ward(1997)}]{Ward}
Ward, W.~R. 1997, Icarus, 126, 261.
\newblock \url{https://www.sciencedirect.com/science/article/pii/S001910359695647X}

\bibitem[{{Welbanks} {et~al.}(2024){Welbanks}, {Bell}, {Beatty}, {Line}, {Ohno}, {Fortney}, {Schlawin}, {Greene}, {Rauscher}, {McGill}, {Murphy}, {Parmentier}, {Tang}, {Edelman}, {Mukherjee}, {Wiser}, {Lagage}, {Dyrek}, \& {Arnold}}]{Welbanks2024}
{Welbanks}, L., {Bell}, T.~J., {Beatty}, T.~G., {et~al.} 2024, \nat, 630, 836

\bibitem[{{Wittrock} {et~al.}(2023){Wittrock}, {Plavchan}, {Cale}, {Barclay}, {Ludwig}, {Schwarz}, {M{\'e}karnia}, {Triaud}, {Abe}, {Suarez}, {Guillot}, {Conti}, {Collins}, {Waite}, {Kielkopf}, {Collins}, {Dreizler}, {El Mufti}, {Feliz}, {Gaidos}, {Geneser}, {Horne}, {Kane}, {Lowrance}, {Martioli}, {Radford}, {Reefe}, {Roccatagliata}, {Shporer}, {Stassun}, {Stockdale}, {Tan}, {Tanner}, \& {Vega}}]{Wittrock}
{Wittrock}, J.~M., {Plavchan}, P.~P., {Cale}, B.~L., {et~al.} 2023, \aj, 166, 232

\bibitem[{{Wong} \& {Lee}(2024)}]{Wong2024}
{Wong}, K.~H., \& {Lee}, M.~H. 2024, \aj, 167, 112

\bibitem[{{Wood} {et~al.}(2023){Wood}, {Mann}, {Barber}, {Bush}, {Kraus}, {Tofflemire}, {Vanderburg}, {Newton}, {Feiden}, {Zhou}, {Bouma}, {Quinn}, {Armstrong}, {Osborn}, {Adibekyan}, {Mena}, {Sousa}, {Gagn{\'e}}, {Fields}, {Milburn}, {Thao}, {Schmidt}, {Gnilka}, {Howell}, {Law}, {Ziegler}, {Brice{\~n}o}, {Ricker}, {Vanderspek}, {Latham}, {Seager}, {Winn}, {Jenkins}, {Schlieder}, {Osborn}, {Twicken}, {Ciardi}, \& {Huang}}]{Wood}
{Wood}, M.~L., {Mann}, A.~W., {Barber}, M.~G., {et~al.} 2023, \aj, 165, 85

\bibitem[{{Wu}(2005)}]{Yanqin2005}
{Wu}, Y. 2005, \apj, 635, 688

\bibitem[{{Wu} {et~al.}(2024{\natexlab{a}}){Wu}, {Chen}, \& {Lin}}]{Wu_Chen_Lin2024}
{Wu}, Y., {Chen}, Y.-X., \& {Lin}, D. N.~C. 2024{\natexlab{a}}, \mnras, 528, L127

\bibitem[{{Wu} {et~al.}(2024{\natexlab{b}}){Wu}, {Malhotra}, \& {Lithwick}}]{Wu2024}
{Wu}, Y., {Malhotra}, R., \& {Lithwick}, Y. 2024{\natexlab{b}}, arXiv e-prints, arXiv:2405.08893

\bibitem[{{Yoder}(1995)}]{Yoder}
{Yoder}, C.~F. 1995, in Global Earth Physics: A Handbook of Physical Constants, ed. T.~J. {Ahrens}, 1

\bibitem[{{Yu} \& {Dai}(2024)}]{Yu_and_Dai}
{Yu}, H., \& {Dai}, F. 2024, \apj, 972, 159

\bibitem[{{Zhang} {et~al.}(2024){Zhang}, {Hu}, {Inglis}, {Dai}, {Bean}, {Knutson}, {Lam}, {Goffo}, \& {Gandolfi}}]{Zhang2024}
{Zhang}, M., {Hu}, R., {Inglis}, J., {et~al.} 2024, \apjl, 961, L44

\end{thebibliography}

\section*{Appendix}
Table:

\begin{longtable}{p{2cm} p{1cm} p{1.5cm} p{1.2cm} p{2cm} p{3.2cm} p{2cm} p{1.5cm}}
\caption{\textbf{Characteristics of Sample Near Resonance}. The uncertainty in \(\log_{10}Q_p^\prime\) represents the standard deviation of \(Q_p^\prime\) values derived from randomly selected parameters (after three effects), as detailed in the main text. \text{$N_P$} denotes the number of planets in the system. \(P_{\text{obs}}\) is the period of the inner planet within the pair. The standard deviation of age from Berger et al is $3.9$. }\\
\label{tab}\\
\toprule
$log_{10}Q_p^\prime$ & $N_P$ & $\Delta_{obs}$  & $P_{obs}$ (days)  & Radius of Planet (R$_{\oplus}$) & Planet Name (inner planet of a pair) &Planet mass ($M_\oplus$)& Age from Berger et al (Gyr) \\
\midrule
\endfirsthead
\caption[]{(Continued)} \\
\toprule
$log_{10}Q_p^\prime$& $N_P$ & $\Delta_{obs}$ & $P_{obs}$ (days)  & Planet Radius (R$_{\oplus}$) & Planet Name (inner planet of a pair) &Planet mass ($M\oplus$)& Age from Berger et al \\
\midrule
\endhead
\midrule
\endfoot
\bottomrule
\endlastfoot

-2.9$\pm$0.43 & 1 & 0.023 & 6.8 & 0.76 & Kepler-431 b & 0.41 & 5.6 \\
-1.2$\pm$0.63 & 1 & 0.010 & 3.6 & 0.40 & Kepler-444 b & 0.045 & 6.1 \\
-1.1$\pm$0.60 & 1 & 0.025 & 2.9 & 0.65 & Kepler-1542 c & 0.23 & 12 \\
-1.0$\pm$0.60 & 3 & 0.023 & 28 & 1.9 & Kepler-341 d & 4.8 & 11 \\
-0.85$\pm$0.48 & 1 & 0.026 & 7.5 & 0.84 & Kepler-450 d & 0.56 & 4.6 \\
-0.53$\pm$0.48 & 1 & 0.013 & 7.4 & 0.74 & Kepler-345 b & 0.36 & 9.6 \\
-0.033$\pm$0.63 & 1 & 0.028 & 5.2 & 1.2 & Kepler-341 b & 1.9 & 6.4 \\
0.075$\pm$0.73 & 1 & 0.013 & 15 & 1.4 & Kepler-128 b & 3.8 & 9.7 \\
0.24$\pm$0.54 & 1 & 0.020 & 7.7 & 2.7 & Kepler-880 b & 6.7 & 9.6 \\
0.24$\pm$0.71 & 1 & 0.023 & 12 & 1.4 & Kepler-386 b & 3.5 & 9.2 \\
0.28$\pm$0.52 & 2 & 0.026 & 9.8 & 2.0 & Kepler-244 c & 5.2 & 11 \\
0.30$\pm$0.57 & 1 & 0.022 & 8.5 & 1.8 & Kepler-331 b & 4.7 & 11 \\
0.31$\pm$0.49 & 1 & 0.023 & 6.5 & 1.0 & Kepler-192 d & 1.1 & 4.8 \\
0.31$\pm$0.73 & 1 & 0.0034 & 23 & 1.1 & Kepler-384 b & 1.6 & 0.36 \\
0.40$\pm$0.53 & 1 & 0.022 & 2.5 & 1.1 & Kepler-327 b & 1.5 & 5.2 \\
0.56$\pm$0.60 & 1 & 0.0100 & 12 & 1.1 & Kepler-59 b & 1.5 & 2.3 \\
0.61$\pm$0.68 & 1 & 0.021 & 12 & 1.0 & Kepler-595 c & 3.3 & 8.4 \\
0.66$\pm$0.55 & 2 & 0.0073 & 7.0 & 1.1 & Kepler-339 c & 1.7 & 4.6 \\
0.75$\pm$0.74 & 1 & 0.018 & 14 & 1.4 & Kepler-127 b & 3.6 & 3.1 \\
0.89$\pm$0.56 & 2 & 0.021 & 3.3 & 1.1 & Kepler-374 c & 1.5 & 2.1 \\
0.92$\pm$0.75 & 1 & 0.025 & 15 & 2.5 & Kepler-340 b & 6.4 & 3.2 \\
0.96$\pm$0.56 & 2 & 0.010 & 8.0 & 1.6 & Kepler-394 b & 4.2 & 5.5 \\
1.2$\pm$0.49 & 1 & 0.0031 & 5.3 & 0.46 & Kepler-102 b & 1.1 & 11 \\
1.2$\pm$0.86 & 2 & 0.0099 & 10 & 1.2 & Kepler-197 c & 2.2 & 0.39 \\
1.2$\pm$0.70 & 3 & 0.0027 & 26 & 2.0 & Kepler-342 c & 5.0 & 6.8 \\
1.3$\pm$0.46 & 1 & 0.023 & 5.7 & 2.1 & Kepler-183 b & 5.3 & 8.2 \\
1.3$\pm$0.51 & 2 & 0.011 & 6.2 & 1.2 & Kepler-169 c & 2.1 & 9.0 \\
1.5$\pm$0.62 & 1 & 0.028 & 29 & 3.9 & Kepler-30 b & 11 & 0.28 \\
1.5$\pm$0.67 & 1 & 0.011 & 10 & 1.8 & Kepler-11 b & 1.9 & 4.6 \\
1.6$\pm$0.77 & 1 & 0.016 & 13 & 3.5 & Kepler-79 b & 11 & 8.0 \\
1.6$\pm$0.61 & 1 & 0.013 & 6.3 & 2.1 & Kepler-120 b & 5.5 & 3.4 \\
1.7$\pm$0.63 & 1 & 0.028 & 3.2 & 1.5 & Kepler-107 b & 3.8 & 5.5 \\
1.8$\pm$0.55 & 1 & 0.0012 & 7.5 & 0.80 & Kepler-1972 b & 2.0 & 1.7 \\
1.9$\pm$0.60 & 1 & 0.025 & 3.4 & 2.0 & Kepler-267 b & 5.1 & 0.86 \\
1.9$\pm$0.63 & 1 & 0.028 & 3.1 & 1.3 & Kepler-181 b & 2.5 & 5.3 \\
1.9$\pm$0.63 & 1 & 0.022 & 21 & 5.5 & Kepler-31 b & 13 & 1.9 \\
2.0$\pm$0.53 & 1 & 0.017 & 5.3 & 2.5 & Kepler-269 b & 6.2 & 5.5 \\
2.1$\pm$0.47 & 1 & 0.020 & 6.2 & 2.7 & Kepler-25 b & 8.7 & 3.2 \\
2.1$\pm$0.62 & 1 & 0.018 & 3.9 & 1.6 & Kepler-226 b & 4.0 & 12 \\
2.1$\pm$0.52 & 2 & 0.0098 & 8.1 & 2.3 & Kepler-24 b & 11 & 3.3 \\
2.2$\pm$0.60 & 1 & 0.028 & 10 & 2.5 & Kepler-29 b & 5.0 & 10.0 \\
2.2$\pm$0.49 & 1 & 0.0048 & 8.3 & 1.8 & Kepler-85 b & 1.8 & 7.0 \\
2.4$\pm$0.65 & 1 & 0.019 & 3.0 & 1.4 & Kepler-272 b & 3.9 & 8.7 \\
2.4$\pm$0.49 & 1 & 0.0085 & 7.1 & 1.6 & Kepler-23 b & 2.6 & 3.2 \\
2.4$\pm$0.40 & 2 & 0.028 & 9.8 & 2.8 & Kepler-83 b & 7.1 & 8.6 \\
2.4$\pm$0.69 & 2 & 0.0069 & 12 & 2.1 & Kepler-254 c & 5.5 & 4.3 \\
2.4$\pm$0.80 & 2 & 0.0091 & 13 & 2.6 & Kepler-176 c & 6.5 & 5.8 \\
2.4$\pm$0.72 & 1 & 0.027 & 2.6 & 1.6 & Kepler-1530 b & 4.3 & 3.0 \\
2.6$\pm$0.72 & 1 & 0.019 & 2.2 & 1.5 & Kepler-326 b & 4.0 & 5.3 \\
2.7$\pm$0.53 & 1 & 0.014 & 4.0 & 1.2 & Kepler-402 b & 2.2 & 8.9 \\
2.7$\pm$0.55 & 1 & 0.014 & 5.9 & 2.0 & Kepler-28 b & 1.6 & 4.4 \\
2.8$\pm$0.76 & 1 & 0.018 & 2.8 & 1.7 & Kepler-221 b & 4.4 & 4.9 \\
2.9$\pm$0.68 & 1 & 0.0016 & 21 & 1.2 & KOI-3503 b & 9.2 & 3.7 \\
3.1$\pm$0.58 & 1 & 0.014 & 5.7 & 3.1 & Kepler-57 b & 25 & 1.8 \\
3.2$\pm$0.64 & 4 & 0.0055 & 28 & 2.4 & Kepler-55 b & 6.1 & 1.7 \\
3.3$\pm$0.61 & 2 & 0.0095 & 5.7 & 1.7 & Kepler-968 c & 4.4 & 1.1 \\
3.3$\pm$0.77 & 2 & 0.016 & 10 & 2.8 & Kepler-58 b & 36 & 4.5 \\
3.4$\pm$0.66 & 1 & 0.012 & 4.8 & 1.9 & Kepler-48 b & 3.9 & 11 \\
3.6$\pm$0.67 & 1 & 0.0065 & 10 & 2.7 & Kepler-385 b & 6.8 & 3.5 \\
3.6$\pm$0.70 & 1 & 0.019 & 11 & 6.5 & Kepler-56 b & 22 & 1.6 \\
3.7$\pm$0.67 & 1 & 0.0036 & 10 & 2.4 & Kepler-307 b & 7.4 & 5.7 \\
3.8$\pm$0.60 & 1 & 0.015 & 2.4 & 1.6 & Kepler-1065 c & 4.2 & 5.9 \\
3.8$\pm$0.73 & 1 & 0.0014 & 14 & 1.9 & KOI-1599.02 & 9.0 & 11 \\
4.0$\pm$0.79 & 2 & 0.022 & 15 & 4.0 & Kepler-27 b & 9.8 & 8.4 \\
4.1$\pm$0.55 & 1 & 0.027 & 3.1 & 1.6 & Kepler-416 d & 4.2 & 3.8 \\
4.3$\pm$0.62 & 2 & 0.0079 & 3.1 & 1.5 & Kepler-80 d & 6.8 & 10 \\
4.3$\pm$0.56 & 2 & 0.010 & 7.2 & 2.6 & Kepler-49 b & 9.8 & 1.2 \\
4.3$\pm$0.54 & 1 & 0.011 & 6.0 & 2.4 & Kepler-81 b & 6.1 & 9.3 \\
4.3$\pm$0.58 & 2 & 0.019 & 2.9 & 1.5 & Kepler-32 e & 4.6 & 3.8 \\
4.5$\pm$0.56 & 2 & 0.013 & 19 & 8.3 & Kepler-9 b & 43 & 3.1 \\
4.6$\pm$0.84 & 1 & 0.0022 & 10 & 0.64 & Kepler-138 b & 0.070 & 0.34 \\
4.8$\pm$0.44 & 1 & 0.0046 & 8.0 & 2.1 & Kepler-54 b & 5.4 & 6.7 \\
5.0$\pm$0.66 & 2 & 0.0056 & 4.7 & 1.2 & Kepler-1669 d & 2.2 & 7.9 \\
5.1$\pm$0.60 & 2 & 0.0074 & 5.5 & 3.6 & Kepler-305 b & 10 & 6.0 \\
5.2$\pm$0.59 & 2 & 0.00040 & 20 & 2.1 & Kepler-372 c & 5.3 & 2.2 \\
6.5$\pm$0.43 & 1 & 0.00022 & 7.1 & 1.7 & Kepler-60 b & 4.2 & 5.4 \\
6.6$\pm$0.62 & 1 & 0.00015 & 7.8 & 1.7 & Kepler-50 b & 4.4 & 7.6 \\
6.7$\pm$0.50 & 2 & 0.024 & 1.2 & 0.78 & Kepler-42 b & 0.44 & 3.4 \\

\end{longtable}

\keywords{exoplanets tides, dynamics}
\end{document}